\documentclass[aps,prb,twocolumn,superscriptaddress,preprintnumbers]{revtex4} 
\usepackage[colorlinks,bookmarks=false,citecolor=blue,linkcolor=red,urlcolor=blue]{hyperref}
\usepackage{epsfig}
\usepackage{tikz}
\usepackage{graphicx}
\usepackage{physics}

\usepackage{dcolumn}
\usepackage{bm}

\usepackage{amsmath,amssymb}

\makeatletter
\renewcommand*\env@matrix[1][*\c@MaxMatrixCols c]{%
  \hskip -\arraycolsep
  \let\@ifnextchar\new@ifnextchar
  \array{#1}}
\makeatother

\usepackage{appendix}

\newcommand{\nn}{\nonumber}

\newcommand{\bea}{\begin{eqnarray}}
\newcommand{\eea}{\end{eqnarray}}         

\begin{document}

\title{The attractive Hubbard model as an $SO(3)$ system of competing phases: \\supersolid order and its thermal melting}

\author{Madhuparna Karmakar}
\affiliation{Department of Physics, Indian Institute of Technology Madras, Chennai 600036, India}
\author{R. Ganesh}
\affiliation{The Institute of Mathematical Sciences, HBNI, C I T Campus, Chennai 600113, India}

\date{\today}

\begin{abstract}
Competition between superconductivity and charge order is a recurring theme in contemporary condensed matter physics. This is quintessentially captured in the attractive Hubbard model, a simple theoretical model where the competition can be directly tuned. In previous studies by the current authors, it has been suggested that the Hubbard model maps to an $SO(3)$ non-linear sigma model, where the phase competition becomes manifest. In this article, we rigorously demonstrate this mapping and 
use it to study thermal disordering of a supersolid. 
Starting with the attractive Hubbard model in the presence of an orbital field, we take the limit of strong coupling where a pseudospin description emerges. The in-plane pseudospin components represent superconducting pairing while the out-of-plane component encodes charge density wave order.  
We obtain an effective spin-$1/2$ Hamiltonian with ferromagnetic in-plane couplings and antiferromagnetic z-z couplings. In addition, the orbital field gives rise to a textured Dzyaloshinskii-Moriya interaction that has the same periodicity as the magnetic unit cell. In order to examine the nature of ordering in this spin model, we consider it in the classical limit. We assume slowly varying fields, leading to the $SO(3)$ non-linear sigma model description.  
As an application of these ideas, we study the nature of ordering using simulated annealing and classical Monte Carlo simulations. The ground state represents a supersolid with coexisting superconductivity and charge order. It can be viewed as a `meron crystal', a regular arrangement of superconducting vortices with charge-ordered cores. 
The overlap of core regions gives rise to 
coherent long-ranged charge order. As the temperature is raised, this charge order is lost via a sharp phase transition in the Ising universality class.
\end{abstract}
\maketitle
\section{Introduction}
Experiments on the underdoped cuprates have fuelled renewed interest in phase competition, with superconductivity competing with charge density wave (CDW) order\cite{Howald2003,ting2004,Wise2008,Gabovich2010,Chang2012,Ghiringhelli2012,Wu2013,LeBoeuf2013,Grissonnanche2014,Nie2015,Machida2015,Gerber2015,Chang2016, Yu2016,Jang2016,Leroux2019}. Studies have highlighted several manifestations of phase competition including ordered vortex cores\cite{Machida2015}, coexistence\cite{Chang2012}, strong impurity response\cite{Leroux2019}, non-monotonic evolution of critical fields\cite{Grissonnanche2014}, etc. Manifestations of phase competition have also been seen in other material families such as transition metal dichalcogenides\cite{Morosan2006,Kusmartseva2009,Kiss2007,Liu2016,Cho2018,Yang2018},  pnictides\cite{Lee2019} and Ba$_{1-x}$K$_x$BiO$_3$\cite{Sleight1975,Cava1988,Sleight2015}.
However, the physics in these materials is obscured by complications such as disorder and incommensurate ordering vectors. We require a simple model system where the effects of phase competition can be first understood in a clean setting. The attractive Hubbard model fits this requirement as a simple system that is amenable to various theoretical approaches. 

A particularly interesting consequence of phase competition occurs in the presence of an orbital field. The superconducting order parameter forms vortices with superconductivity suppressed within the core region of each vortex. This allows for the competing CDW to arise locally. When the vortex density is large, the overlap between neighbouring vortex cores leads to coherent long-ranged CDW order\cite{Wu2013,KarmakarPRB2017}. This leads to a `supersolid' phase that has coexisting superconductivity and CDW orders. Supersolidity has remained elusive in experiments despite intense studies in several contexts\cite{Boninsegni2012}. Overlap of ordered vortex cores provides a new mechanism that could allow for robust and verifiable supersolidity. 
In previous studies by the current authors, this mechanism has been demonstrated in the attractive Hubbard model using a mean-field approach\cite{KarmakarPRB2017}. Remarkably, the results were consistent with an $SO(3)$ field theory for competing phases. Based on this observation, it was conjectured that these two models were equivalent. This conjecture was supported by further studies\cite{KarmakarJPSJ2017,Saran2019}. In this article, we establish this equivalence by way of a rigorous mapping. In the process, we find interesting results concerning a `meron crystal', its stability to thermal fluctuations and melting.

One of our key results is the derivation of an $SO(3)$ non-linear sigma model. This has strong similarities to the well-known $SO(5)$ theory proposed in the context of the cuprates. The $SO(5)$ theory is written in terms of a five component vector field: two corresponding to superconductivity and three to antiferromagnetism\cite{Demler2004}. Although several consequences were worked out\cite{Arovas1997,Hu2002}, the model has not 
received support from experiments on the cuprates. In addition, it remains a phenomenological construct as no microscopic origin has been demonstrated. Here, we present a similar, but simpler, theory with a three-component order parameter. We start with a precise microscopic model and derive an effective $SO(3)$ theory. This is potentially directly testable in experiments, with several proposals for realizing the attractive Hubbard model in ultracold atomic gases\cite{Tarruell2018}.

\section{The Attractive Hubbard model at strong coupling}
We consider particles on a square lattice with nearest- and next-nearest-neighbour hopping, $t$ and $t'$. 
When two particles are on the same site, they lower the energy of the system due to an attractive interaction. Due to Pauli exclusion, this is only possible if they carry opposite spin. This leads to the Hamiltonian
\bea
\nn H &=& -t \sum_{\langle ij \rangle,\sigma}  e^{i\theta_{ij}} c_{i,\sigma}^\dagger c_{j,\sigma} 
-t'\sum_{\ll ij \gg,\sigma} e^{i\theta_{ij}} c_{i,\sigma}^\dagger c_{j,\sigma} + h.c.\\
& -& \mu \sum_{i,\sigma} n_{i,\sigma} 
- U \sum_{i} (n_{i,\uparrow}-1/2)(n_{i,\downarrow}-1/2).
\eea
We have introduced Peierls' phases in the hopping amplitudes, $\theta_{ij}$, that originate from a uniform orbital magnetic field. 
This can be viewed as an Aharonov-Bohm phase accrued by the particle as it hops from one site to the next. We have introduced a chemical potential $\mu$ to fix particle density. In the rest of this article, we will restrict our attention to half-filling and use $t'$ as a handle to tune phase competition. 

The competition between superconductivity and CDW orders arises from a special $SU(2)$ symmetry, first pointed out by C. N. Yang\cite{Yang1989,Yang1990,Zhang1990}. It requires three conditions: (a) a bipartite lattice with hopping between opposite sublattices, (b) absence of the orbital field, with all $\theta_{ij}=0$, and (c) the density being fixed at half-filling. The square lattice with purely nearest neighbour hopping ($t'=0$) meets these requirements. At half-filling, it has perfect degeneracy between superconductivity and CDW orders. Upon introducing a next-nearest neighbour $t'$ hopping, this degeneracy is lost with a superconducting ground state. Nevertheless, CDW order remains as low-lying competitor with the energy cost scaling as $\sim t'^2$\cite{Ganeshthesis}.
 
The $SU(2)$ symmetry is best seen in the strong coupling limit of the Hubbard model ($U\gg t$). Previous studies have shown that the Hubbard model maps to a $S=1/2$ pseudospin XXZ model\cite{Burkov2008}. This can be further mapped to a Heisenberg model using a sublattice-dependent spin rotation. With a non-zero $t'$, the CDW state manifests as a low-lying `roton' excitation in the spin wave spectrum\cite{Ganesh2009,Yunomae2009}. In this article, we derive the pseudospin Hamiltonian in the presence of an orbital field. Going further, we derive a coarse-grained field theory from the spin model. We study the role of thermal fluctuations by investigating the pseudospin model using Monte Carlo simulations.

\section{Strong coupling pseudospin model}
\begin{figure}
\includegraphics[width=2in]{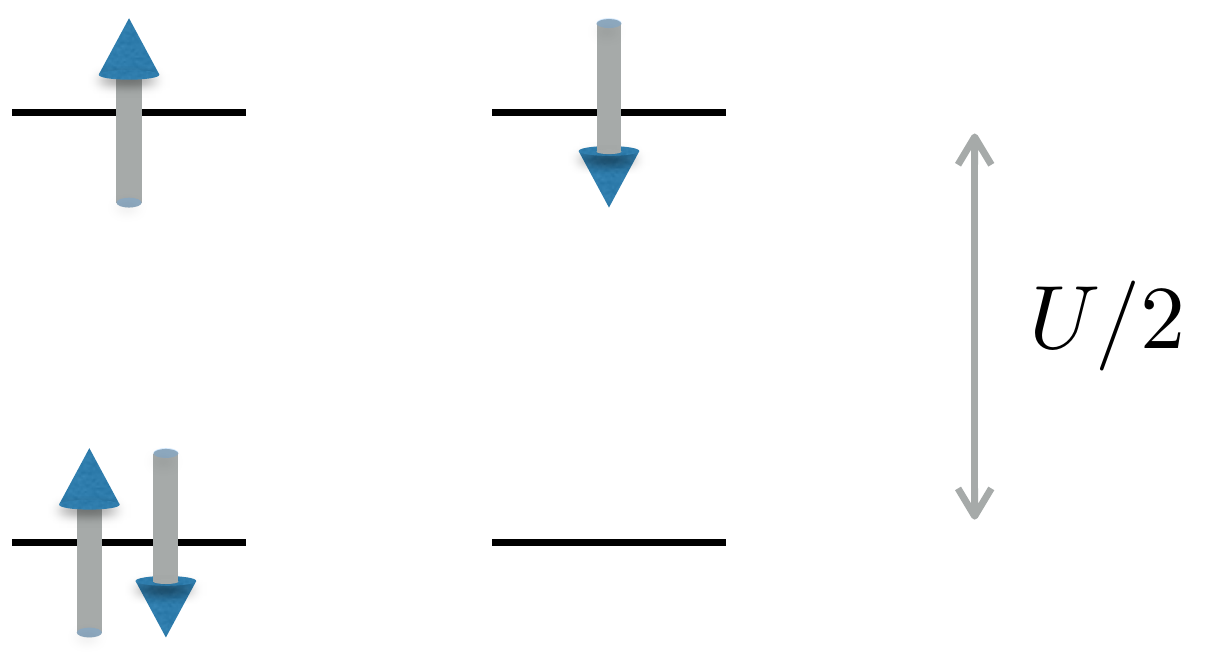}
\caption{Single site Hilbert space. Empty and doubly occupied states are approximately equal in energy. Singly occupied states have an energy cost of $U$. }
\label{fig.singlesite}
\end{figure}

We consider the strong coupling limit of the model with $U\gg t,t'$, following the superexchange scheme that has been presented in Refs.~\onlinecite{Burkov2008,Ganesh2009}. If we only keep this dominant $U$-term in the Hamiltonian, the sites decouple from one another, leaving a purely on-site problem. The spectrum for the single site problem is shown in Fig.~\ref{fig.singlesite}. The energy of the singly occupied states is $-\mu+U/4$ as can be seen from the Hamiltonian above. The energy of the empty state is $-U/4$. Likewise, the energy of the doubly occupied state is $-2\mu-U/4$. 

At half-filling, the empty and doubly occupied sites must have the same energy so that they are occupied with the same probability. To ensure this, we set $\mu= 0$. 
The spectrum splits into two pairs of states as shown in Fig.~\ref{fig.singlesite}. The empty and doubly occupied states have lower energy, while the singly occupied states have higher energy. The energy difference between the pairs of states is $U/2$. The hopping terms in the Hamiltonian act as small perturbations on these states. Their effect is seen at second order in perturbation theory where they couple two sites at a time. To see this explicitly, we consider a two-site problem next.

\subsection{Two site problem}
We consider two sites labelled $A$ and $B$. 
They may represent nearest neighbours or next-nearest neighbours on the square lattice. Apart from the dominant on-site terms, the Hamiltonian contains inter-site hopping terms,
\begin{equation}
H_{AB}^{hop} = t_{AB}\sum_\sigma c_{B,\sigma}^\dagger c_{A,\sigma}  + t_{AB}^*\sum_\sigma c_{A,\sigma}^\dagger c_{B,\sigma}.
\end{equation}
The hopping amplitude $t_{AB}$ can be complex with its phase given by the Peierl's substitution scheme. We reexpress it as $t_{AB}=\tau_{AB}e^{i\theta_{AB}}$.

\begin{figure}
\includegraphics[width=3.5in]{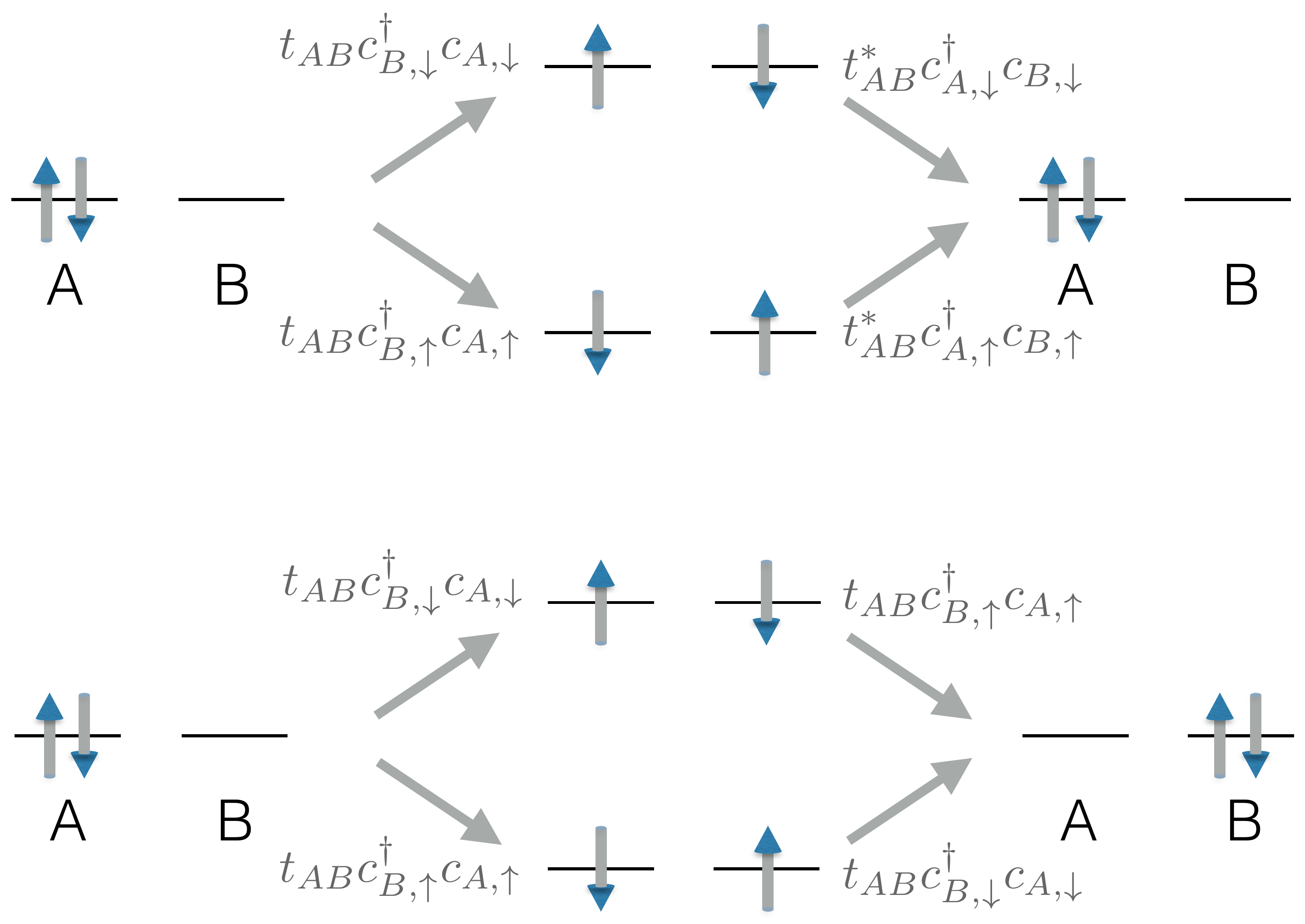}
\caption{Superexchange pathways: The two panels depict processes that involve two sites -- one doubly occupied and one empty. In the top panel, the initial and final states are the same, i.e., the pseudospin at each site is preserved. In the bottom panel, the initially empty state becomes doubly occupied and vice versa. This represents an exchange of pseudospins.   }
\label{fig.pathways}
\end{figure}

We now consider the low energy Hilbert space of the two-site problem. 
We introduce a pseudospin notation for the low energy states on a given site. We denote the empty state as pseudospin-down ($\vert \!\!\Downarrow \rangle$) and the doubly occupied state as pseudospin-up ($\vert \!\!\Uparrow \rangle$). In the two-site Hilbert space, we have four low energy states with each site having pseudospin-up or -down. As the hopping term takes us out of this subspace, we treat it within perturbation theory. Indeed, there are second order processes that connect low energy states, as shown in 
Fig.~\ref{fig.pathways}. In each path in the figure, the intermediate state has two singly occupied states. As a result, it has an energy cost given by $2\times U/2 = U$.

The two-site states with parallel pseudospins (both empty or both doubly occupied) are unaffected within second order. In states with antiparallel spins, we find two processes: one that preserves pseudospins and one that exchanges them. We obtain the following Hamiltonian
\begin{eqnarray}
H_{\mathcal{O}(t^2)} = \Psi_{AB}^\dagger H_{4\times4}\Psi_{AB},
\end{eqnarray}
where $\Psi_{AB} = \big(\langle  \Uparrow_A \Uparrow_B\!\vert , 
\langle  \Uparrow_A \Downarrow_B\!\vert, 
\langle   \Downarrow_A \Uparrow_B\!\vert, 
\langle  \Downarrow_A \Downarrow_B\!\vert 
\big)^T$. The $4\times4$ Hamiltonian matrix is given by
\begin{eqnarray}
H_{4\times4}=\left(\begin{array}{cccc}
-U/2 & 0 & 0 & 0 \\
0 &  C & D & 0 \\
0 & D^* & C & 0 \\
0 & 0 & 0 & -U/2-4\mu
\end{array}\right),
\end{eqnarray}
with $C =-U/2+(-2 \tau_{AB}^2/U)$ and $D = -2\tau_{AB}^2 e^{-2i\theta_{AB}}/U$. 
The diagonal terms have a contribution from second order perturbation theory, in the form of $C$. In this term, the two hopping processes contribute with opposite phases that cancel out. In contrast, the phases add in the off-diagonal $D$ term, imbuing it with a phase of $2\theta_{AB}$. 

We now add a constant shift of $U/2 + \tau_{AB}^2/U$ along the diagonals. The resulting Hamiltonian can be expressed in terms of an effective exchange coupling, $J=4\tau_{AB}^2/U$, 
\begin{eqnarray}
H_{4\times4}\sim\left(\begin{array}{cccc}
J/4 & 0 & 0 & 0 \\
0 & -J/4& -(J/2) e^{-2i\theta_{AB}} & 0 \\
0 & -(J/2) e^{2i\theta_{AB}} & (-J/4) & 0 \\
0 & 0 & 0 & J/4
\end{array}\right).
\label{eq.H4x4}
\end{eqnarray}
This matrix has a simple interpretation in terms of spin operators. It can be written as
\begin{equation}
H_{AB} = J\left[ \hat{S}_{A}^z \hat{S}_{B}^z -\frac{1}{2}\left\{ e^{-2i\theta_{AB}}  \hat{S}_{A}^+ \hat{S}_{B}^- + e^{2i\theta_{AB}}  \hat{S}_{A}^- \hat{S}_{B}^+ \right\}  \right], 
\label{eq.SHamiltonian}
\end{equation} 
where $\hat{S}$ are pseudospin-$1/2$ operators. 
This can be rewritten as follows,
\begin{eqnarray}
\nn H_{AB} = J\big[ \hat{S}_{A}^z \hat{S}_{B}^z &-& \cos(2\theta_{AB}) \left\{ \hat{S}_{A}^x \hat{S}_{B}^x + \hat{S}_{A}^y \hat{S}_{B}^y \right\}  \\
&+&  \sin(2\theta_{AB})  \left\{ \hat{S}_{A}^x \hat{S}_{B}^y - \hat{S}_{A}^y \hat{S}_{B}^x \right\}
  \big]. 
\label{eq.SHamiltonianb}
\end{eqnarray} 
The term proportional to $\sin(2\theta_{AB})$ can be expressed as $\vec{D}\cdot (\vec{\hat{S}}_A \times \vec{\hat{S}}_B)$, where $\vec{D} = \{0,0,J\sin(2\theta_{AB})\}$ -- a Dzyaloshinskii-Moriya interaction\cite{Dzyaloshinsky1958,Moriya1960}. The term proportional to $\cos(2\theta_{AB})$ represents an XY-like exchange coupling between in-plane components.
Note that the coupling constant, $J$, depends on the hopping strength on the bond. For example, it will have different strengths along nearest and next-nearest bonds. 

\subsection{Pseudospin model on the lattice}
\label{ssec.pseudospin}
We have defined a pseudospin operator on each site. Its $z$-component represents the local CDW order parameter. To see this, we note that a site with pseudospin-up is doubly occupied with positive deviation from half-filling, whereas a site with pseudospin-down is empty with negative deviation.
A state with maximal CDW order corresponds to an alternating arrangement of empty and doubly-occupied sites. This corresponds to an antiferromagnetic pseudospin arrangement with moments pointing alternately along $\pm \hat{z}$. On the other hand, the in-plane pseudospin components represent superconductivity. More precisely, the $x$ and $y$ components represent the real and imaginary parts of the pairing order parameter. This can be seen from the $SU(2)$ pseudospin operators, $\hat{S}_i^x \equiv \frac{1}{2}\{c_{i,\downarrow}^\dagger c_{i,\uparrow}^\dagger + c_{i,\uparrow} c_{i,\downarrow} \}$ and $\hat{S}_i^y \equiv \frac{1}{2i} \{ c_{i,\downarrow}^\dagger c_{i,\uparrow}^\dagger - c_{i,\uparrow} c_{i,\downarrow} \}$. Superconductivity is signalled by non-zero expectation values for these operators.

Extending the two-particle effective Hamiltonian to the lattice, we arrive at a square lattice spin problem with the Hamiltonian
\bea
\nn H =  \sum_{\langle ij \rangle, \ll ij \gg} J_{ij} \Big[  \hat{S}_i^z \hat{S}_j^z &+& \gamma_{ij} (\hat{S}_i^x \hat{S}_j^y + \hat{S}_i^y \hat{S}_j^x) \\
&+& \delta_{ij} \hat{z} \cdot (\vec{\hat{S}}_i \times \vec{\hat{S}}_j)\Big] .
\label{eq.Hspin}
\eea
The coupling strengths are given by $J = 4t^2/U$ and $J' = 4t'^2/U$ on nearest and next-nearest neighbours respectively. The bond-dependent exchange and Dzyalonshinskii-Moriya coefficients are given $\gamma_{ij} = \cos(2\theta_{ij})$ and $\delta_{ij} = \sin(2\theta_{ij})$. The latter two depend on $\theta_{ij}$, the Peierls' phase associated with the bond $(ij)$. 

In the initial Hubbard model, the Peierls' phases encode a uniform orbital magnetic field. They are given by $\theta_{ij}= (e/\hbar)\int_i^j \mathbf{A} \cdot \mathbf{dl}$, where $\mathbf{A}$ is the magnetic vector potential. Several studies have explored ways to realize this physical setup in ultracold atomic gases\cite{Jaksch2003,Schweikhard2004,Lin2009,An2017,Aidelsburger2013,Miyake2013}.
 In a superconductor, strictly speaking, the orbital field must be self-consistently determined using Maxwell's equations. For the sake of simplicity, we assume a uniform orbital magnetic field below. This is a reasonable assumption in strongly type-II superconductors. 
The results discussed in Sec.~\ref{sec.so3} below hold regardless of this assumption.

As the vector potential is not unique, neither is the assignment of Peierls' phases. If the vector potential is altered by a gauge transformation, this can be absorbed into the in-plane spin components by a suitable redefinition. This can be seen from Eq.~\ref{eq.Hspin}, where in-plane pseudospin components couple to the $\theta_{ij}$'s while the $z$ components do not. This is consistent with the identification of the in-plane components with the superconducting order parameter. In this sense, the effective model of Eq.~\ref{eq.Hspin} should \textit{not} be thought of as a true spin model, as the in-plane spin components are gauge-dependent quantities. 

\begin{figure*}
\includegraphics[width=2\columnwidth]{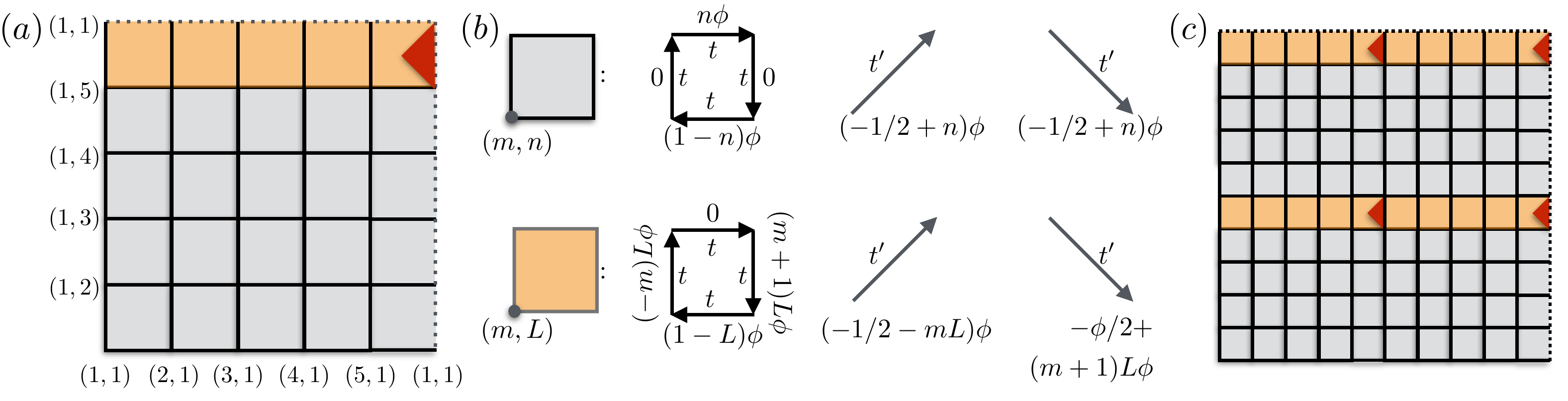}
\caption{Peierls phases in the lattice. A $5\times5$ magnetic unit cell is shown in (a), with two types of cells: yellow and grey. The Peierls' phases are assigned as shown in (b). The net flux through the $5\times5$ block corresponding to a single flux quantum. The red region at top right corner of (a) contains the Dirac string, an anomalous flux that adds an unobservable Aharonov-Bohm phase of $2\pi$.   
(c) A $20\times20$ lattice where Peierls' phases are assigned to form 4 magnetic unit cells.}
\label{fig.Peierls}
\end{figure*}

Traditionally, spin models are studied on finite lattices using periodic boundary conditions. Taking such an approach to Eq.~\ref{eq.Hspin} leads to some fundamental issues. We first note that the Peierls' phases necessarily contain singularities. To see this, we note that the square lattice forms a closed surface (a torus) due to periodic boundary conditions. A net flux through the lattice corresponds to having a magnetic monopole charge inside the torus. As argued by Dirac\cite{Dirac1931}, the vector potential cannot be smoothly defined on a surface enclosing a magnetic monopole. It necessarily includes flux tubes, called Dirac strings, that impart an Aharanov-Bohm phase of $2\pi$. The number of Dirac strings is equal to the number of flux quanta that pierce the lattice. It follows that the Peierls' phases ($\theta_{ij}$'s) cannot have the same periodicity of the underlying lattice. They must necessarily form a large unit cell. 
The smallest possible unit cell corresponds to the area that contains a single Dirac string, i.e., the area carrying a single flux quantum. In other words, it is the `magnetic unit cell'. 
One such phase assignment is shown in Fig.~\ref{fig.Peierls}. This leads to the Hamiltonian in Eq.~\ref{eq.Hspin} with translational symmetry such that the unit cell is the same as the magnetic unit cell.

Using periodic boundaries has a second important consequence. Considering a charged particle on a surface enclosing a magnetic monopole, Dirac showed that its wavefunction cannot be defined in a smooth manner\cite{Dirac1931,Wu1976}. In the system at hand, the superconducting order parameter cannot be smoothly defined on the torus. This can be seen as a consequence of having a non-zero number of vortices and no compensating anti-vortices. In the spin model, the in-plane spin components will not vary smoothly on the square lattice. They will invariably contain singularities or jumps. This serves as an additional caveat in viewing Eq.~\ref{eq.Hspin} as a spin problem.

\section{The $SO(3)$ effective field theory}
\label{sec.so3}
In the previous section, we arrived at an effective pseudospin description, assuming half-filling and $U \gg t,t'$. We now show that this pseudo-spin problem gives rise to a non-linear sigma model in the low energy limit. 

We begin with the pseudospin Hamiltonian of Eq.~\ref{eq.Hspin} on an infinite square lattice. Promoting the spins to the classical limit, we have a lattice problem with three-dimensional vector moments. We make two further assumptions: (a) at low energies (low temperatures), the spin configurations are `smooth' with small gradients, and (b) with a weak orbital field, the Peierls' phase on each bond is small. We now note that Eq.~\ref{eq.Hspin} has \textit{antiferromagnetic} $z-z$ couplings between nearest neighbours. In contrast, the in-plane couplings are \textit{ferromagnetic} (for small $\theta_{ij}$'s). This indicates that, in low-energy configurations, the spins are of the form, $S\times(\Delta_x, \Delta_y, (-1)^\mathbf{r} \rho)$, where $\Delta_x$, $\Delta_y$ and $\rho$ are slowly varying quantities satisfying $\Delta_x^2 + \Delta_y^2 + \rho^2 = 1$. Here, $S$ denotes the spin length. We henceforth set $S=1$ for simplicity.
The z-component carries a rapid oscillation given by $(-1)^\mathbf{r}$, which varies in a checkerboard fashion on the square lattice. 
As we expect $\Delta_x$, $\Delta_y$ and $\rho$ to vary smoothly on the scale of the lattice constant, we elevate them to slowly varying fields, $\Delta_x(\mathbf{r})$, $\Delta_y(\mathbf{r})$ and $\rho(\mathbf{r})$ respectively. Note that the spatially-averaged $z$-moment vanishes. This corresponds to the assumption of half-filling, as the $z$-moment represents the local deviation from half-filling. 
We now calculate the contribution from each term in Eq.~\ref{eq.Hspin} within the language of coarse-grained fields.

\subsection{CDW terms}
\label{ssec.cdw}
We first consider the $z-z$ couplings in Eq.~\ref{eq.Hspin} that resemble those of an Ising model on the square lattice,
\begin{eqnarray}
H^{zz} = \sum_{m,n} E_{m,n}^{zz},
\end{eqnarray}
where $(m,n)$ represents a site on the square lattice. The contribution from each site is given by
\begin{eqnarray}
\nonumber E_{m,n}^{zz} &=& \frac{J}{2} S_{m,n}^z \Big[  S_{m+1,n}^z + S_{m-1,n}^z + S_{m,n+1}^z + S_{m,n-1}^z \Big] \\
\nonumber &+& \frac{J'}{2} S_{m,n}^z \Big[  S_{m+1,n+1}^z + S_{m-1,n-1}^z \\
&~& ~~~~~~~~+S_{m-1,n+1}^z + S_{m+1,n-1}^z \Big].
\end{eqnarray}
The factors of $1/2$ have been added to avoid double counting. 
We reinterpret this energy density in terms of the coarse-grained $\rho(\mathbf{r})$ field. We use $\hat{S}_{m,n}^z \approx (-1)^{m+n} \rho(\mathbf{r}_{m,n})$, where $(m,n)$ denotes a site of the square lattice. 
As with the standard Ising model, we elevate the summation over $(m,n)$ to an integral and reexpress the integrand using $\rho(\mathbf{r})$ and its derivatives. We obtain
\begin{eqnarray}
H^{zz} \approx \int dx dy ~  \Big[  -a_\rho \rho^2(\mathbf{r}) + \chi_\rho 
\vert \vec{\nabla} \rho(\mathbf{r}) \vert^2 \Big].
\label{eq.LGrho}
\end{eqnarray}
where $a_\rho = 2\{J-J' \}$ and $\chi_\rho = \frac{(J -2J'  )\ell^2}{2}$. Here, $\ell$ denotes the lattice constant of the square lattice. In these two coefficients, the $J$ and $J'$ appear with opposite sign. This stems from the rapidly oscillating $(-1)^{m+n}$ factor that takes the opposite (same) sign on (next-) nearest neighbours. In addition, their relative amplitudes are different in $a_\rho$ and $\chi_\rho$, i.e., we have $a_\rho \sim (J-J')$ while $\chi_\rho\sim (J -2J')$. This difference arises from the differing bond lengths for nearest ($\ell$) and next-nearest ($\sqrt{2}\ell$) neighbours. 

\subsection{Superconducting terms}
We now consider the in-plane terms in the pseudospin Hamiltonian. In order to get a better understanding, we first take the vector potential to be zero, i.e., we ignore the Peierls' phases. This leads to a two-component spin model on the square lattice with ferromagnetic XY couplings. We have
\begin{eqnarray}
H^{xy}_{\vec{A}=0} = -J \sum_{\langle ij \rangle} \vec{S}_{i,\parallel} \cdot \vec{S}_{j,\parallel} - J' \sum_{\ll ij \gg}
\vec{S}_{i,\parallel} \cdot \vec{S}_{j,\parallel}, 
\label{eq.Htb}
\end{eqnarray} 
where $\vec{S}_{i,\parallel} = (S_i^x,S_i^y)$. Taking the in-plane components to be described by the slowly-varying fields $\Delta_x(\mathbf{r})$ and $\Delta_y(\mathbf{r})$, we obtain the field theory of an XY ferromagnet,
\begin{eqnarray}
\nonumber H^{xy}_{\vec{A}=0} \approx \int dxdy~  \Big[ -a_\Delta \vert \vec{\Delta} (\mathbf{r})\vert^2 
&+&\chi_\Delta  \{ \vec{\nabla} \Delta_x \cdot \vec{\nabla} \Delta_x +\\
&~& \vec{\nabla} \Delta_y \cdot \vec{\nabla} \Delta_y
\}
\Big],
\label{eq.Hxy}
\end{eqnarray} 
where $\vec{\Delta} \equiv (\Delta_x,\Delta_y)$, $a_\Delta = 2\{ J + J'\}$ and $\chi_\Delta = \frac{(J+2J')\ell^2}{2}$.
This can be seen in direct analogy with the CDW term above, by replacing $\rho$ with $\Delta_{x/y}$. Unlike the CDW terms, the $J$ and $J'$ contributions have the same sign here. 

We now draw an analogy to the problem of a free particle in two-dimensional space. We take its wavefunction to be $\Delta (\mathbf{r}) = \Delta_x (\mathbf{r}) + i \Delta_y (\mathbf{r})$. Taking its mass to be $1/2\chi_\Delta$ and assuming a constant potential $(-a_\Delta)$, its Hamiltonian is given by $\hat{\mathcal{H}} = \chi_\Delta \hat{p}^2 + a_\Delta$. The expectation value of the Hamiltonian is then precisely given by Eq.~\ref{eq.Hxy}. A discrete form of this Hamiltonian can be constructed using a tight-binding-like approach. Discretizing the space as a square mesh with sites denoted by $(m,n)$, we take $\Delta(\mathbf{r}) \rightarrow S_{m,n}^x+i S_{m,n}^y$. This leads to the Hamiltonian in Eq.~\ref{eq.Htb}. This analogy provides a simple interpretation for in-plane terms in the Hamiltonian: the superconducting order parameter represents the wavefunction of a free particle (the Cooper pair). 

We now introduce an orbital magnetic field. By comparing the Eqs.~\ref{eq.Htb}, \ref{eq.Hspin} and \ref{eq.SHamiltonian}, we see that the orbital field enters as Peierls' phases in a tight binding Hamiltonian. The superconducting wavefunction couples to the vector potential as a charged particle with charge $2e$. 
It can immediately be deduced that the vector potential enters Eq.~\ref{eq.Htb} via the well known minimal coupling prescription,
\begin{eqnarray}
\nonumber H^{xy} \approx \int dxdy~  \Big[ -a_\Delta \vert \vec{\Delta} (\mathbf{r})\vert^2 
&+&\chi_\Delta  \{\vec{\mathcal{D}} \Delta_x \cdot \vec{\mathcal{D}} \Delta_x +\\
&~&\vec{\mathcal{D}} \Delta_y \cdot \vec{\mathcal{D}} \Delta_y
\}
\Big],~~~
\label{eq.Hxy_A}
\end{eqnarray} 
where $\vec{\mathcal{D}} \equiv \vec{\nabla} + i\frac{2e}{\hbar} \vec{A} $. Note that the charge here is $2e$, that of a Cooper pair.
Indeed, we find the same result by a systematic analysis of the in-plane terms. The orbital field modifies Eq.~\ref{eq.Htb} to give
\begin{eqnarray}
\nonumber &~&\!\!H^{xy} =  \sum_{m,n} E_{m,n}^{xy};\\
\nonumber &~&\!\! E_{m,n}^{xy} = \frac{J}{2} \sum_{\vec{\delta}}  \Big[ 
\cos (\mathcal{A}_{m,n,\vec{\delta}})  \vec{S}_{m,n}^\parallel \cdot \vec{S}_{m+\delta_x,n+\delta_y}^\parallel  \\
\nonumber &-& \sin (\mathcal{A}_{m,n,\vec{\delta} }) \{ S_{m,n}^x S_{m+\delta_x,n+\delta_y}^y 
 -  S_{m,n}^y S_{m+\delta_x,n+\delta_y}^x \}
\Big] \\
\nonumber &+&\frac{J'}{2} \sum_{\vec{\eta}}  \Big[ 
\cos (\mathcal{A}_{m,n,\vec{\eta}})  \vec{S}_{m,n}^\parallel \cdot \vec{S}_{m+\eta_x,n+\eta_y}^\parallel  \\
 &-& \sin (\mathcal{A}_{m,n,\vec{\eta}}) \{ S_{m,n}^x S_{m+\eta_x,n+\eta_y}^y 
 -  S_{m,n}^y S_{m+\eta_x,n+\eta_y}^x \}
\Big].~~~~~~
\end{eqnarray}
where $\vec{\delta}$ and $\vec{\eta}$ sum over the nearest and next-nearest neighbour vectors respectively. Here, $\mathcal{A}$'s denote Peierls' phases, e.g., $\mathcal{A}_{m,n,\delta} = \frac{2e}{\hbar}\int_{(m,n)}^{(m+\delta_x,n+\delta_y)} \mathbf{A} \cdot \mathbf{dl}$.
Assuming slow variations in the $\Delta$'s and small values of the Peierls' angles, we precisely recover Eq.~\ref{eq.Hxy_A}. This follows the usual derivation of the long-wavelength minimal-coupling Hamiltonian from a tight binding model with Peierls' phases. 

\subsection{The non-linear sigma model}
Combining the CDW and superconducting contributions from Eqs.~\ref{eq.LGrho} and \ref{eq.Hxy_A}, we obtain the Hamiltonian density in terms of coarse-grained fields,
\begin{eqnarray}
\nonumber \mathcal{H} &=&   -a_\rho \rho^2(\mathbf{r})   -a_\Delta  \vert \Delta (\mathbf{r}) \vert^2 \\
&+&  \chi_\rho 
\vert \vec{\nabla} \rho (\mathbf{r}) \vert^2 + \chi_\Delta  \Big\{ \vec{\mathcal{D}} \Delta(\mathbf{r}) \Big\}^* \cdot \Big\{ \vec{\mathcal{D}}\Delta(\mathbf{r}) \Big\}.
\end{eqnarray}
We have combined $\Delta_x$ and $\Delta_y$ into a single complex field, $\Delta(\mathbf{r}) \equiv \Delta_x (\mathbf{r}) + i \Delta_y (\mathbf{r})$. The fields $\rho(\mathbf{r})$ and $\Delta(\mathbf{r})$ are not independent, as they must necessarily satisfy a uniform length constraint, $\rho^2 + \vert \Delta \vert^2 = 1$. The coefficients are given by $a_\rho = 2\{J-J' \}$, $\chi_\rho = \frac{(J -2J'  )a^2}{2}$, $a_\Delta = 2\{ J + J'\}$ and $\chi_\Delta = \frac{(J+2J')a^2}{2}$. Rescaling allows us to write a simpler form,
\begin{eqnarray}
\nonumber \mathcal{H} &\approx&    -\vert \Delta (\mathbf{r}) \vert^2 - (1-\epsilon) \rho^2(\mathbf{r}) \\
 &+& \chi  \Big\{ \vec{\mathcal{D}} \Delta(\mathbf{r}) \Big\}^* \cdot \Big\{ \vec{\mathcal{D}}\Delta(\mathbf{r}) \Big\} + \chi (1-\xi) 
\vert \vec{\nabla} \rho (\mathbf{r}) \vert^2,~~~
\label{eq.NLSM}
\end{eqnarray}
where $\epsilon = 2 J'/(J+J')$, $\chi = \frac{a^2}{4}\{1+\epsilon/2\}$ 
and $\xi = 4J'/(J+2J')$. If $t'$ is small in the microscopic Hubbard problem, we have $\epsilon, \xi \sim (t'/t)^2$ with $\xi \approx 2\epsilon$. Here, $\epsilon$ and $\xi$ reflect the anisotropy between superconductivity and CDW order. The $SO(3)$ character of this model can be seen by setting $t'$ and the orbital field to zero. In this limit, Eq.~\ref{eq.NLSM} reduces to the Hamiltonian density of a symmetric Heisenberg ferromagnet. When weak anisotropies are introduced, the physics retains signatures of the proximate $SO(3)$ point.

This form is closely related to the previously conjectured model in Ref.~\onlinecite{KarmakarPRB2017}, where the anisotropy in the gradient term was ignored (i.e., $\xi$ was set to zero). Nevertheless, this does not lead to any qualitative change in the physics of phase competition. We see this below in the nature of the ground state.   

\section{Simulating the non-linear sigma model}
We have shown that the attractive Hubbard model reduces to an $SO(3)$ non-linear sigma model. Using this equivalence, we seek to study its physics in the presence of an orbital field. The energy of the system is given by the Hamiltonian density of Eq.~\ref{eq.NLSM}. The ground state can be found by minimizing the energy, subject to the uniform length constraint ($\rho^2  +\vert \Delta \vert^2 = 1$). However, minimizing Eq.~\ref{eq.NLSM} on the infinite two-dimensional plane is a non-trivial task. Likewise, thermal properties of the non-linear sigma model can be found by averaging over configurations with a suitable Boltzmann weight. Once again, this is a difficult task on the infinite plane.

We approach this problem by reversing the arguments put forward in the previous sections. We now view Eq.~\ref{eq.Hspin}, the pseudospin model on the square lattice, as a regularization of the non-linear sigma model in Eq.~\ref{eq.NLSM}. We will study the pseudospin model on finite lattices with periodic boundary conditions and look for results that remain consistent upon increasing system size. 
This opens the door to well established techniques from the field of magnetism. In particular, we use simulated annealing to find the ground state of Eq.~\ref{eq.Hspin}. We will interpret the result in terms of the smooth fields of the non-linear sigma model. We will then study the role of thermal fluctuations using classical Monte Carlo simulations. 

The pseudospin model of Eq.~\ref{eq.Hspin} is defined on the square lattice. As explained in Sec.~\ref{ssec.pseudospin} above, the Hamiltonian depends on the choice of the Peierls' phases. We present results using the scheme depicted in Fig.~\ref{fig.Peierls}. We assume a $12\times 12$ magnetic unit cell so that the Peierls' phases do not vary too rapidly from one bond to the next. We consider a lattice composed of an $n\times n$ array of magnetic unit cells, giving rise to a $12n\times 12n$ lattice with periodic boundaries. We approach the thermodynamic limit by increasing $n$. 

\subsection{Supersolidity in the ground state}

\begin{figure*}
\includegraphics[width=2\columnwidth]{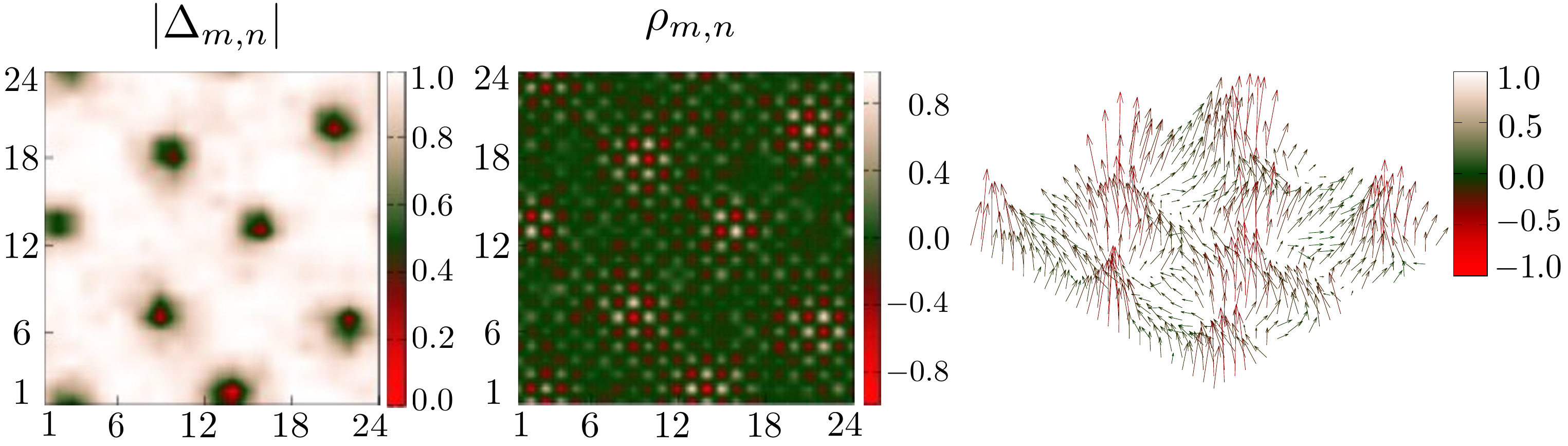}
\caption{Lowest energy state obtained from simulated annealing on a $24\times24$ lattice, with $t'=0.2t$. Left: The superconducting amplitude, showing a regular arrangement of vortices. Centre: The CDW order parameter, showing CDW ordering emerging at vortex cores and percolating throughout. Right: The resulting pseudospin texture. In the $z$ component of the pseudospin, we have removed the rapidly oscillating factor of $(-1)^{m+n}$ to allow for a clear visualization. The arrow colour has been set to reflect the local $z$-component. }
\label{fig.simal}
\end{figure*}
To find the lowest energy state, we perform simulated annealing of the pseudospin model.  We use two types of single-site moves: Metropolis and microcanonical (overrelaxation). At each site, we find the effective field that arises from the neighbouring moments. The Metropolis move corresponds to changing the inclination with respect to the effective field. The microcanonical move rotates the spin about the effective field so as to preserve the energy.

The lowest energy state found from simulated annealing is shown in Fig.~\ref{fig.simal}. We have used a $24\times 24$ lattice containing four $12\times12$ magnetic unit cells. The net magnetic flux through the lattice thus corresponds to four flux quanta. 
At each site, we interpret the in-plane components of the pseudospin as the superconducting order parameter. From Fig.~\ref{fig.simal}(left), we see that the superconducting amplitude vanishes at regularly spaced points, indicating a vortex lattice. The number of vortices is eight, with two vortices for each flux quantum. We have defined a flux quantum with respect to the charge $e$ of the particle hopping on the lattice. As a Cooper pair has charge $2e$, we find two vortices for each flux quantum. 

The competition with CDW order is clearly seen in Fig.~\ref{fig.simal}(centre) which shows the $z$-component of the spins in the ground state. We see strong CDW order appearing in each vortex core. The CDW order percolates through the inter-vortex space and covers the entire lattice. This leads to a `meron crystal' as shown in Fig.~\ref{fig.simal}(right). Here, we plot $(S_{m,n}^x,S_{m,n}^y,(-1)^{m+n}S_{m,n}^z )$ vs. $(m,n)$, i.e., position on the lattice. This conveys the variation of the pseudospin orientation in space. We have removed a rapidly oscillating phase in the z-component of the pseudospin (see discussion in Sec.~\ref{ssec.cdw} above). 
Each superconducting vortex takes the form of a `meron' in the pseudospin. The in-plane components wind by $2\pi$ as we move around the vortex. Within the core region, an out-of-plane component develops to preserve the spin length. Due to overlap between adjacent merons, the out-of-plane component is non-zero everywhere. It has the same sign at all sites, indicating coherent CDW order.

This picture is consistent with the results of Ref.~\onlinecite{KarmakarPRB2017} where the Hubbard model was directly studied using Bogoliubov-deGennes mean field simulations. In particular, the low energy state here represents a `supersolid'. It has well-defined superconducting order that is reflected in the formation of a vortex lattice. At the same time, it has long-ranged CDW order.

\subsection{Classical Monte Carlo simulations}
We have established that ground state of the non-linear sigma model in Eq.~\ref{eq.NLSM} is a supersolid with coexisting superconductivity and CDW order. The superconductivity sector encapsulates an additional layer of ordering in the form of a vortex lattice with discrete translational symmetry. Upon increasing the temperature, we may see multiple phase transitions where these orders melt independently. To study thermal fluctuations, we study the pseudospin model of Eq.~\ref{eq.Hspin} using classical Monte Carlo simulations. We use single-site Metropolis and microcanonical (overrelaxation) moves. We start from a random initial configuration on an $L\times L$ lattice at high temperature and progressively decrease the temperature. At each temperature, we perform 8 $\times$10$^6$ sweeps, each with $L^{2}$ single-site moves, with the ratio of Metropolis to microcanonical fixed at 4:3. The first 2 $\times$ 10$^5$ moves are discarded to allow for equilibration. 

\begin{figure*}
\includegraphics[width=2\columnwidth]{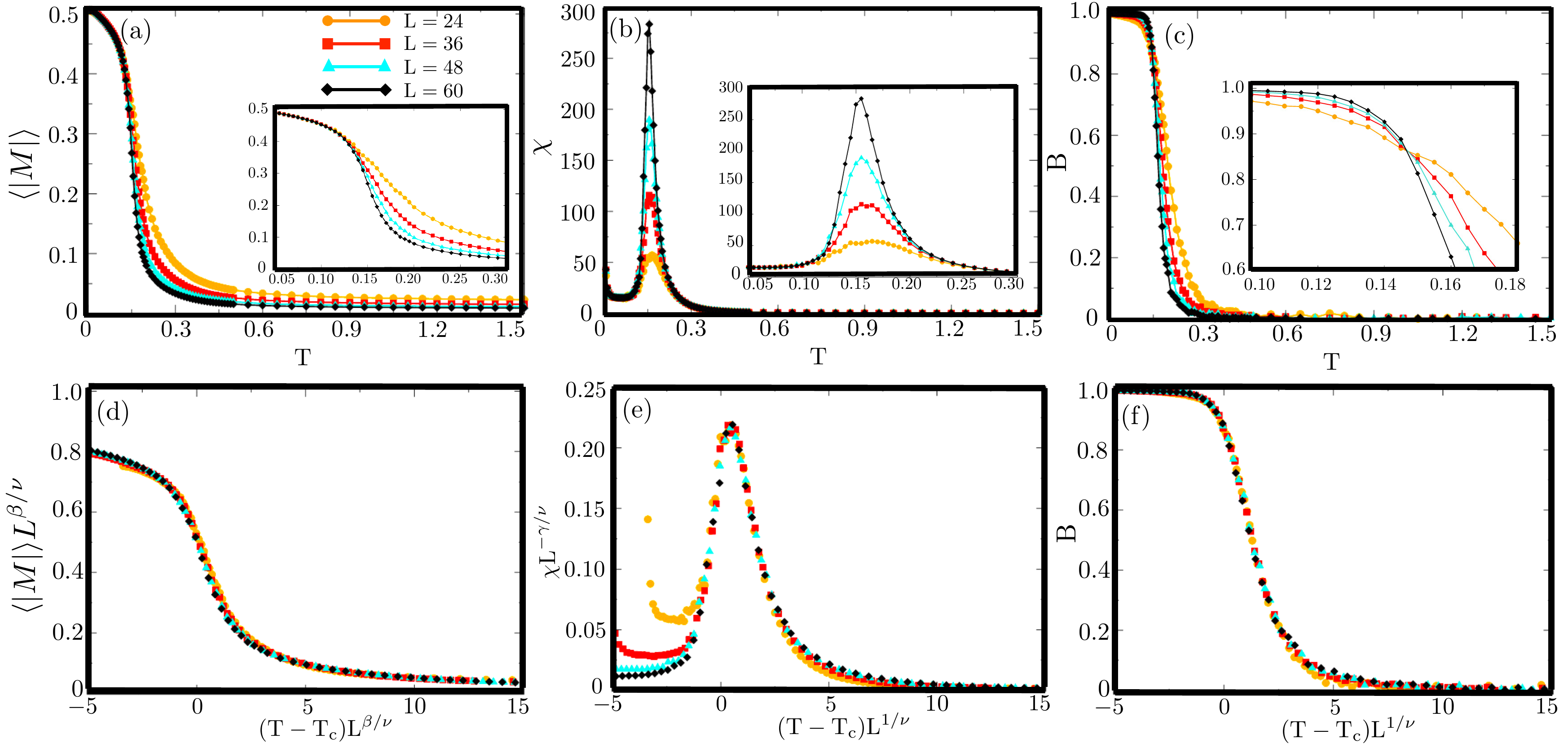}
\caption{Ising phase transition as seen from classical Monte Carlo simulations with $t'=0.15t$. Panels (a), (b) and (c) show the temperature dependence of the magnetization, susceptibility and Binder cumulant respectively. The insets are zoomed-in plots of the same data over a narrow temperature range. 
Panels (d), (e) and (f) show the the same data after scaling using the critical exponents of the 2D Ising model. 
}
\label{fig.ising}
\end{figure*}
We first discuss the thermal evolution of the CDW order. In the non-linear sigma model of Eq.~\ref{eq.NLSM}, the CDW order parameter shows an Ising-like character with the energy being invariant under $\rho(\mathbf{r}) \rightarrow -\rho(\mathbf{r})$. This originates from the Hubbard model where the CDW order represents a checkerboard-like modulation in density. The Ising degree of freedom corresponds to choosing one of the two sublattices as that with higher density. In the pseudospin model of Eq.~\ref{eq.Hspin}, the CDW order parameter is the staggered $z$-magnetization, given by $M =  \frac{1}{L^2}\sum_{m,n} (-1)^{m+n} S_{m,n}^z$, where $L$ is the linear system size. We define the corresponding susceptibility and Binder cumulant as $\chi = \frac{L^2}{T}(\langle M ^2 \rangle - \langle \vert M \vert \rangle^2)$ and $B =\frac{3}{2} -\frac{1}{2} \langle  M^4 \rangle / (\langle  M^2 \rangle)^2  $. Here, $\langle .\rangle$ represents averaging over Monte Carlo  configurations. The coefficients in the Binder cumulant are designed so as to (a) vanish in the high temperature paramagnetic phase and (b) approach unity in the case of maximal CDW ordering. 

The temperature dependence of the order parameter is shown in Fig.~\ref{fig.ising}(a). We find a profile that is typical of an Ising magnet. Starting from zero at high temperatures, it approaches a non-zero value at low temperatures. Unlike the standard Ising magnet, the magnetization in the zero-temperature-limit is not unity. This can be understood from the ground state configuration in Fig.~\ref{fig.simal}. The CDW order is not uniform; rather, it has maximal intensity at vortex cores and weak order at inter-vortex positions. Nevertheless, we see a clear indication of an Ising-like phase transition. Fig.~\ref{fig.ising}(a) shows the order parameter for various system sizes with $L=12 n$, where $n=2,3,4,5$. We choose $L$ to be multiples of $12$ so that we can construct the pseudospin Hamiltonian using a $12\times12$ magnetic unit cell. The flux density is the same for all system sizes. We find further evidence for a phase transition in the form of a peak in susceptibility as shown in Fig.~\ref{fig.ising}(b). The peak height grows with system size as expected.

To determine the precise location of the CDW phase transition, we examine the Binder cumulant for various system sizes, shown in Fig.~\ref{fig.ising}(c). We find a crossing at $T_c \approx 0.145 \pm 0.001$. We surmise that this transition belongs to the universality class of the 2D Ising model. To verify this, we perform a scaling analysis of the data. In Fig.~\ref{fig.ising}(d-f), we plot the rescaled order parameter, susceptibility and Binder cumulant vs. reduced temperature (using $T_c$ as obtained from the Binder cumulant crossing). We find good scaling collapse using the well known critical exponents of the 2D Ising model\cite{Baxter2007}, viz., $\nu = 1$, $\beta = 1/8$ and $\gamma=7/4$. Based on this finding, we assert that CDW order vanishes via a continuous phase transition in the 2D Ising universality class.

\begin{figure*}
\includegraphics[width=2\columnwidth]{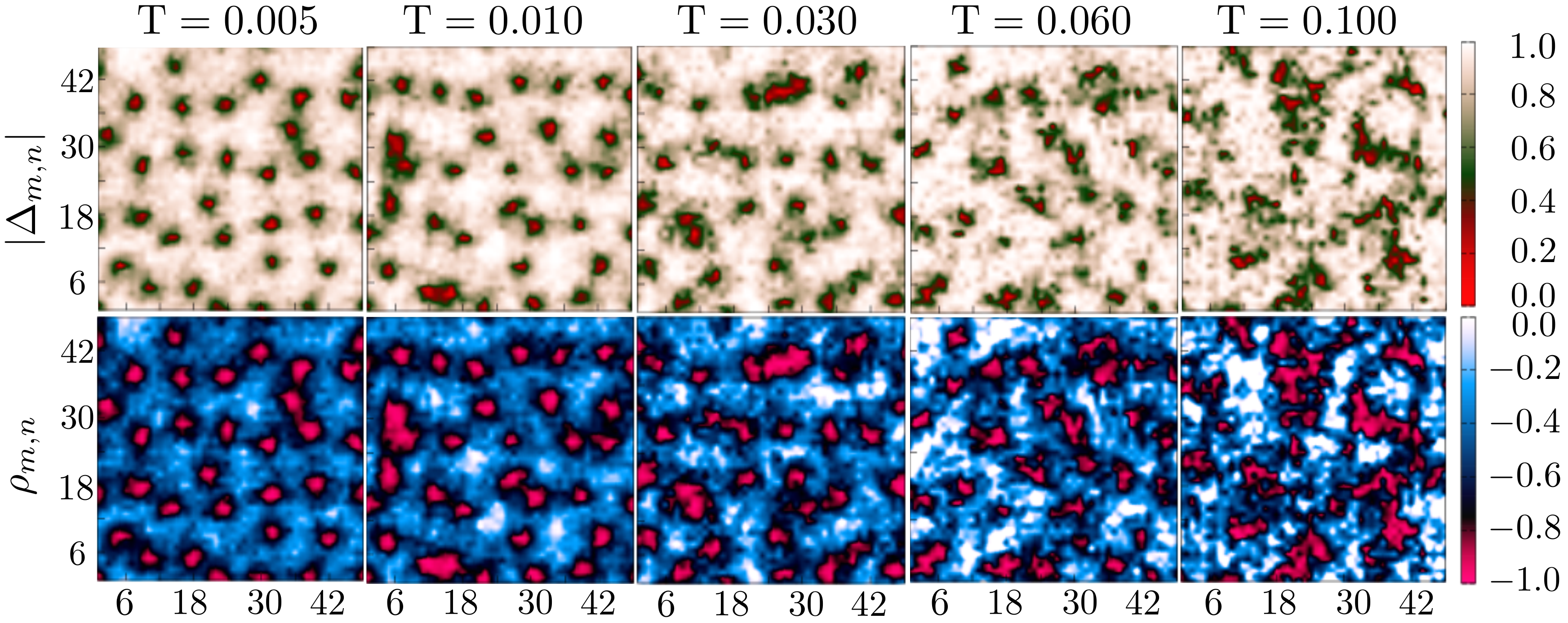}
\caption{Typical configurations seen in Monte Carlo runs at different temperatures with $t'=0.15t$. The panels on top show the superconducting amplitude, i.e., the length of the in-plane projections of pseudospin vectors. The panels in the bottom show the CDW order parameter, i.e., the $z$-component of the pseudospins.   
}
\label{fig.snapshots}
\end{figure*}

We next discuss thermal evolution of the superconducting order. We do not find a distinct phase transition within our Monte Carlo scheme. We believe this is due to technical limitations, as discussed below. Nevertheless, a qualitative understanding can be gained by examining typical configurations extracted from the Monte Carlo simulations, shown in Fig.~\ref{fig.snapshots}. At low temperatures, we see a vortex lattice, albeit with small distortions. The distortions increase with increasing temperature. Beyond $T\sim 0.03$, the vortex lattice is lost as some vortices come close to one another and essentially fuse. At this point, we may view the system as being deep inside a vortex liquid phase. 

We believe that a vortex melting transition occurs at $T\lesssim 0.005$. However, this is not discernible in our simulations as the spins do not relax adequately at low temperatures. Indeed, we do not find a perfect vortex lattice even at the lowest temperatures as some distortions persist (see Fig.~\ref{fig.simal}). This could be a consequence of our single-site update scheme. At a more subtle level, this could be a consequence of the gauge structure. Our system with periodic boundaries cannot support a smoothly varying superconducting field. As discussed in Sec.~\ref{ssec.pseudospin} above, it must necessarily contain singularities or jumps. On account of these discontinuities, a single-site update scheme may not be able to explore the space of all low energy configurations.   

\section{Discussion}
We have presented a study of phase competition in the attractive Hubbard model at strong coupling. We demonstrate a mapping to a pseudospin problem and further onto an $SO(3)$ field theory. This brings out phase competition as an inherent feature of this model. It also reveals an interesting role for an orbital magnetic field as it induces vortices in the superconducting order, but with CDW-ordered cores. Indeed, we find a supersolid ground state with phase coexistence arising from vortex-core-overlap. In the language of spins, we find a meron crystal -- an emergent crystalline phase with a mesoscopic lattice scale, analogous to the well known skyrmion crystal phase. With increasing temperature, superconductivity and CDW orders melt independently with a sharp Ising phase transition in the CDW sector.

Our results bear similarities with disordering transitions in other systems with coexisting orders. We mention two examples from the field of magnetism: (a) The square $J_1-J_2$ antiferromagnet with $J_2 > J_1/2$ breaks $O(3)\otimes\mathbb{Z}_2$ symmetry, where the $\mathbb{Z}_2$ character corresponds to a choice between vertical and horizontal stripes\cite{Chandra1990}. While the $O(3)$ rotational symmetry is restored at an infinitesimal temperature, the $\mathbb{Z}_2$ order persists up until a critical temperature where it is lost via an Ising transition. (b) 
The triangular lattice XY antiferromagnet breaks $\mathbb{Z}_2 \otimes U(1)$ symmetry in the ground state, where the $\mathbb{Z}_2$ character corresponds to a local chirality degree of freedom. The $\mathbb{Z}_2$ order is lost via an Ising transition\cite{Miyashita1984}. In the context of the attractive Hubbard model, we have presented an effective field theory for competing orders. This could be used to potentially develop a renormalization group scheme to understand the physics of disordering. For example, the Ising transition temperature can be lowered by increasing $t'$, i.e., the energy cost of the CDW phase. At a critical value of $t'$, the Ising transition will compete with the vortex lattice melting transition. This can potentially give rise to an interesting combined melting transition.

The pseudospin model derived in Sec.~\ref{ssec.pseudospin} above is essentially a quantum model with $S=1/2$ moments. We have studied this model in the classical limit, taking into account thermal fluctuations. An interesting future direction is to investigate the role of quantum fluctuations. In analogy with the thermal state immediately below the Ising transition, quantum fluctuations may disrupt superconductivity while preserving CDW order. Such a state would represent a `pairing liquid' in analogy with a spin liquid. The pairing liquid offers two advantages over typical spin liquid models: (i) it has an additional tuning handle in the form of an orbital field, (ii) fluctuations of the pairing liquid are intrinsically coupled to the CDW order parameter due to the non-linear uniform length constraint. This offers a new route to probe fluctuations in the liquid phase. These issues may be explored within a quantum treatment of the pseudospin model.

\bibliographystyle{apsrev4-1}
\bibliography{HubbardSO3}

\begin{thebibliography}{54}%
\makeatletter
\providecommand \@ifxundefined [1]{%
 \@ifx{#1\undefined}
}%
\providecommand \@ifnum [1]{%
 \ifnum #1\expandafter \@firstoftwo
 \else \expandafter \@secondoftwo
 \fi
}%
\providecommand \@ifx [1]{%
 \ifx #1\expandafter \@firstoftwo
 \else \expandafter \@secondoftwo
 \fi
}%
\providecommand \natexlab [1]{#1}%
\providecommand \enquote  [1]{``#1''}%
\providecommand \bibnamefont  [1]{#1}%
\providecommand \bibfnamefont [1]{#1}%
\providecommand \citenamefont [1]{#1}%
\providecommand \href@noop [0]{\@secondoftwo}%
\providecommand \href [0]{\begingroup \@sanitize@url \@href}%
\providecommand \@href[1]{\@@startlink{#1}\@@href}%
\providecommand \@@href[1]{\endgroup#1\@@endlink}%
\providecommand \@sanitize@url [0]{\catcode `\\12\catcode `\$12\catcode
  `\&12\catcode `\#12\catcode `\^12\catcode `\_12\catcode `\%12\relax}%
\providecommand \@@startlink[1]{}%
\providecommand \@@endlink[0]{}%
\providecommand \url  [0]{\begingroup\@sanitize@url \@url }%
\providecommand \@url [1]{\endgroup\@href {#1}{\urlprefix }}%
\providecommand \urlprefix  [0]{URL }%
\providecommand \Eprint [0]{\href }%
\providecommand \doibase [0]{http://dx.doi.org/}%
\providecommand \selectlanguage [0]{\@gobble}%
\providecommand \bibinfo  [0]{\@secondoftwo}%
\providecommand \bibfield  [0]{\@secondoftwo}%
\providecommand \translation [1]{[#1]}%
\providecommand \BibitemOpen [0]{}%
\providecommand \bibitemStop [0]{}%
\providecommand \bibitemNoStop [0]{.\EOS\space}%
\providecommand \EOS [0]{\spacefactor3000\relax}%
\providecommand \BibitemShut  [1]{\csname bibitem#1\endcsname}%
\let\auto@bib@innerbib\@empty
\bibitem [{\citenamefont {Howald}\ \emph {et~al.}(2003)\citenamefont {Howald},
  \citenamefont {Eisaki}, \citenamefont {Kaneko}, \citenamefont {Greven},\ and\
  \citenamefont {Kapitulnik}}]{Howald2003}%
  \BibitemOpen
  \bibfield  {author} {\bibinfo {author} {\bibfnamefont {C.}~\bibnamefont
  {Howald}}, \bibinfo {author} {\bibfnamefont {H.}~\bibnamefont {Eisaki}},
  \bibinfo {author} {\bibfnamefont {N.}~\bibnamefont {Kaneko}}, \bibinfo
  {author} {\bibfnamefont {M.}~\bibnamefont {Greven}}, \ and\ \bibinfo {author}
  {\bibfnamefont {A.}~\bibnamefont {Kapitulnik}},\ }\href {\doibase
  10.1103/PhysRevB.67.014533} {\bibfield  {journal} {\bibinfo  {journal} {Phys.
  Rev. B}\ }\textbf {\bibinfo {volume} {67}},\ \bibinfo {pages} {014533}
  (\bibinfo {year} {2003})}\BibitemShut {NoStop}%
\bibitem [{\citenamefont {Chen}\ and\ \citenamefont {Ting}(2005)}]{ting2004}%
  \BibitemOpen
  \bibfield  {author} {\bibinfo {author} {\bibfnamefont {H.-Y.}\ \bibnamefont
  {Chen}}\ and\ \bibinfo {author} {\bibfnamefont {C.~S.}\ \bibnamefont
  {Ting}},\ }\href {\doibase 10.1103/PhysRevB.71.220510} {\bibfield  {journal}
  {\bibinfo  {journal} {Phys. Rev. B}\ }\textbf {\bibinfo {volume} {71}},\
  \bibinfo {pages} {220510} (\bibinfo {year} {2005})}\BibitemShut {NoStop}%
\bibitem [{\citenamefont {Wise}\ \emph {et~al.}(2008)\citenamefont {Wise},
  \citenamefont {Boyer}, \citenamefont {Chatterjee}, \citenamefont {Kondo},
  \citenamefont {Takeuchi}, \citenamefont {Ikuta}, \citenamefont {Wang},\ and\
  \citenamefont {Hudson}}]{Wise2008}%
  \BibitemOpen
  \bibfield  {author} {\bibinfo {author} {\bibfnamefont {W.~D.}\ \bibnamefont
  {Wise}}, \bibinfo {author} {\bibfnamefont {M.~C.}\ \bibnamefont {Boyer}},
  \bibinfo {author} {\bibfnamefont {K.}~\bibnamefont {Chatterjee}}, \bibinfo
  {author} {\bibfnamefont {T.}~\bibnamefont {Kondo}}, \bibinfo {author}
  {\bibfnamefont {T.}~\bibnamefont {Takeuchi}}, \bibinfo {author}
  {\bibfnamefont {H.}~\bibnamefont {Ikuta}}, \bibinfo {author} {\bibfnamefont
  {Y.}~\bibnamefont {Wang}}, \ and\ \bibinfo {author} {\bibfnamefont {E.~W.}\
  \bibnamefont {Hudson}},\ }\href {https://doi.org/10.1038/nphys1021}
  {\bibfield  {journal} {\bibinfo  {journal} {Nature Physics}\ }\textbf
  {\bibinfo {volume} {4}},\ \bibinfo {pages} {696 EP } (\bibinfo {year}
  {2008})}\BibitemShut {NoStop}%
\bibitem [{\citenamefont {Gabovich}\ \emph {et~al.}(2010)\citenamefont
  {Gabovich}, \citenamefont {Voitenko}, \citenamefont {Ekino}, \citenamefont
  {Li}, \citenamefont {Szymczak},\ and\ \citenamefont {Pekala}}]{Gabovich2010}%
  \BibitemOpen
  \bibfield  {author} {\bibinfo {author} {\bibfnamefont {A.~M.}\ \bibnamefont
  {Gabovich}}, \bibinfo {author} {\bibfnamefont {A.~I.}\ \bibnamefont
  {Voitenko}}, \bibinfo {author} {\bibfnamefont {T.}~\bibnamefont {Ekino}},
  \bibinfo {author} {\bibfnamefont {M.~S.}\ \bibnamefont {Li}}, \bibinfo
  {author} {\bibfnamefont {H.}~\bibnamefont {Szymczak}}, \ and\ \bibinfo
  {author} {\bibfnamefont {M.}~\bibnamefont {Pekala}},\ }\href
  {http://dx.doi.org/10.1155/2010/681070} {\bibfield  {journal} {\bibinfo
  {journal} {Advances in Condensed Matter Physics}\ }\textbf {\bibinfo {volume}
  {2010}},\ \bibinfo {pages} {40} (\bibinfo {year} {2010})}\BibitemShut
  {NoStop}%
\bibitem [{\citenamefont {Chang}\ \emph {et~al.}(2012)\citenamefont {Chang},
  \citenamefont {Blackburn}, \citenamefont {Holmes}, \citenamefont
  {Christensen}, \citenamefont {Larsen}, \citenamefont {Mesot}, \citenamefont
  {Liang}, \citenamefont {Bonn}, \citenamefont {Hardy}, \citenamefont
  {Watenphul}, \citenamefont {Zimmermann}, \citenamefont {Forgan},\ and\
  \citenamefont {Hayden}}]{Chang2012}%
  \BibitemOpen
  \bibfield  {author} {\bibinfo {author} {\bibfnamefont {J.}~\bibnamefont
  {Chang}}, \bibinfo {author} {\bibfnamefont {E.}~\bibnamefont {Blackburn}},
  \bibinfo {author} {\bibfnamefont {A.~T.}\ \bibnamefont {Holmes}}, \bibinfo
  {author} {\bibfnamefont {N.~B.}\ \bibnamefont {Christensen}}, \bibinfo
  {author} {\bibfnamefont {J.}~\bibnamefont {Larsen}}, \bibinfo {author}
  {\bibfnamefont {J.}~\bibnamefont {Mesot}}, \bibinfo {author} {\bibfnamefont
  {R.}~\bibnamefont {Liang}}, \bibinfo {author} {\bibfnamefont {D.~A.}\
  \bibnamefont {Bonn}}, \bibinfo {author} {\bibfnamefont {W.~N.}\ \bibnamefont
  {Hardy}}, \bibinfo {author} {\bibfnamefont {A.}~\bibnamefont {Watenphul}},
  \bibinfo {author} {\bibfnamefont {M.~v.}\ \bibnamefont {Zimmermann}},
  \bibinfo {author} {\bibfnamefont {E.~M.}\ \bibnamefont {Forgan}}, \ and\
  \bibinfo {author} {\bibfnamefont {S.~M.}\ \bibnamefont {Hayden}},\ }\href
  {http://dx.doi.org/10.1038/nphys2456} {\bibfield  {journal} {\bibinfo
  {journal} {Nat Phys}\ }\textbf {\bibinfo {volume} {8}},\ \bibinfo {pages}
  {871} (\bibinfo {year} {2012})}\BibitemShut {NoStop}%
\bibitem [{\citenamefont {Ghiringhelli}\ \emph {et~al.}(2012)\citenamefont
  {Ghiringhelli}, \citenamefont {Le~Tacon}, \citenamefont {Minola},
  \citenamefont {Blanco-Canosa}, \citenamefont {Mazzoli}, \citenamefont
  {Brookes}, \citenamefont {De~Luca}, \citenamefont {Frano}, \citenamefont
  {Hawthorn}, \citenamefont {He}, \citenamefont {Loew}, \citenamefont {Sala},
  \citenamefont {Peets}, \citenamefont {Salluzzo}, \citenamefont {Schierle},
  \citenamefont {Sutarto}, \citenamefont {Sawatzky}, \citenamefont {Weschke},
  \citenamefont {Keimer},\ and\ \citenamefont {Braicovich}}]{Ghiringhelli2012}%
  \BibitemOpen
  \bibfield  {author} {\bibinfo {author} {\bibfnamefont {G.}~\bibnamefont
  {Ghiringhelli}}, \bibinfo {author} {\bibfnamefont {M.}~\bibnamefont
  {Le~Tacon}}, \bibinfo {author} {\bibfnamefont {M.}~\bibnamefont {Minola}},
  \bibinfo {author} {\bibfnamefont {S.}~\bibnamefont {Blanco-Canosa}}, \bibinfo
  {author} {\bibfnamefont {C.}~\bibnamefont {Mazzoli}}, \bibinfo {author}
  {\bibfnamefont {N.~B.}\ \bibnamefont {Brookes}}, \bibinfo {author}
  {\bibfnamefont {G.~M.}\ \bibnamefont {De~Luca}}, \bibinfo {author}
  {\bibfnamefont {A.}~\bibnamefont {Frano}}, \bibinfo {author} {\bibfnamefont
  {D.~G.}\ \bibnamefont {Hawthorn}}, \bibinfo {author} {\bibfnamefont
  {F.}~\bibnamefont {He}}, \bibinfo {author} {\bibfnamefont {T.}~\bibnamefont
  {Loew}}, \bibinfo {author} {\bibfnamefont {M.~M.}\ \bibnamefont {Sala}},
  \bibinfo {author} {\bibfnamefont {D.~C.}\ \bibnamefont {Peets}}, \bibinfo
  {author} {\bibfnamefont {M.}~\bibnamefont {Salluzzo}}, \bibinfo {author}
  {\bibfnamefont {E.}~\bibnamefont {Schierle}}, \bibinfo {author}
  {\bibfnamefont {R.}~\bibnamefont {Sutarto}}, \bibinfo {author} {\bibfnamefont
  {G.~A.}\ \bibnamefont {Sawatzky}}, \bibinfo {author} {\bibfnamefont
  {E.}~\bibnamefont {Weschke}}, \bibinfo {author} {\bibfnamefont
  {B.}~\bibnamefont {Keimer}}, \ and\ \bibinfo {author} {\bibfnamefont
  {L.}~\bibnamefont {Braicovich}},\ }\href {\doibase 10.1126/science.1223532}
  {\bibfield  {journal} {\bibinfo  {journal} {Science}\ }\textbf {\bibinfo
  {volume} {337}},\ \bibinfo {pages} {821} (\bibinfo {year}
  {2012})}\BibitemShut {NoStop}%
\bibitem [{\citenamefont {Wu}\ \emph {et~al.}(2013)\citenamefont {Wu},
  \citenamefont {Mayaffre}, \citenamefont {Kr{\"a}mer}, \citenamefont
  {Horvati{\'c}}, \citenamefont {Berthier}, \citenamefont {Kuhns},
  \citenamefont {Reyes}, \citenamefont {Liang}, \citenamefont {Hardy},
  \citenamefont {Bonn},\ and\ \citenamefont {Julien}}]{Wu2013}%
  \BibitemOpen
  \bibfield  {author} {\bibinfo {author} {\bibfnamefont {T.}~\bibnamefont
  {Wu}}, \bibinfo {author} {\bibfnamefont {H.}~\bibnamefont {Mayaffre}},
  \bibinfo {author} {\bibfnamefont {S.}~\bibnamefont {Kr{\"a}mer}}, \bibinfo
  {author} {\bibfnamefont {M.}~\bibnamefont {Horvati{\'c}}}, \bibinfo {author}
  {\bibfnamefont {C.}~\bibnamefont {Berthier}}, \bibinfo {author}
  {\bibfnamefont {P.~L.}\ \bibnamefont {Kuhns}}, \bibinfo {author}
  {\bibfnamefont {A.~P.}\ \bibnamefont {Reyes}}, \bibinfo {author}
  {\bibfnamefont {R.}~\bibnamefont {Liang}}, \bibinfo {author} {\bibfnamefont
  {W.~N.}\ \bibnamefont {Hardy}}, \bibinfo {author} {\bibfnamefont {D.~A.}\
  \bibnamefont {Bonn}}, \ and\ \bibinfo {author} {\bibfnamefont {M.-H.}\
  \bibnamefont {Julien}},\ }\href {http://dx.doi.org/10.1038/ncomms3113}
  {\bibfield  {journal} {\bibinfo  {journal} {Nature Communications}\ }\textbf
  {\bibinfo {volume} {4}},\ \bibinfo {pages} {2113 EP } (\bibinfo {year}
  {2013})}\BibitemShut {NoStop}%
\bibitem [{\citenamefont {LeBoeuf}\ \emph {et~al.}(2013)\citenamefont
  {LeBoeuf}, \citenamefont {Kramer}, \citenamefont {Hardy}, \citenamefont
  {Liang}, \citenamefont {Bonn},\ and\ \citenamefont {Proust}}]{LeBoeuf2013}%
  \BibitemOpen
  \bibfield  {author} {\bibinfo {author} {\bibfnamefont {D.}~\bibnamefont
  {LeBoeuf}}, \bibinfo {author} {\bibfnamefont {S.}~\bibnamefont {Kramer}},
  \bibinfo {author} {\bibfnamefont {W.~N.}\ \bibnamefont {Hardy}}, \bibinfo
  {author} {\bibfnamefont {R.}~\bibnamefont {Liang}}, \bibinfo {author}
  {\bibfnamefont {D.~A.}\ \bibnamefont {Bonn}}, \ and\ \bibinfo {author}
  {\bibfnamefont {C.}~\bibnamefont {Proust}},\ }\href
  {http://dx.doi.org/10.1038/nphys2502} {\bibfield  {journal} {\bibinfo
  {journal} {Nat Phys}\ }\textbf {\bibinfo {volume} {9}},\ \bibinfo {pages}
  {79} (\bibinfo {year} {2013})}\BibitemShut {NoStop}%
\bibitem [{\citenamefont {Grissonnanche}\ \emph {et~al.}(2014)\citenamefont
  {Grissonnanche}, \citenamefont {Cyr-Choini{\`e}re}, \citenamefont
  {Lalibert{\'e}}, \citenamefont {Ren{\'e}de~Cotret}, \citenamefont
  {Juneau-Fecteau}, \citenamefont {Dufour-Beaus{\'e}jour}, \citenamefont
  {Delage}, \citenamefont {LeBoeuf}, \citenamefont {Chang}, \citenamefont
  {Ramshaw}, \citenamefont {Bonn}, \citenamefont {Hardy}, \citenamefont
  {Liang}, \citenamefont {Adachi}, \citenamefont {Hussey}, \citenamefont
  {Vignolle}, \citenamefont {Proust}, \citenamefont {Sutherland}, \citenamefont
  {Kr{\"a}mer}, \citenamefont {Park}, \citenamefont {Graf}, \citenamefont
  {Doiron-Leyraud},\ and\ \citenamefont {Taillefer}}]{Grissonnanche2014}%
  \BibitemOpen
  \bibfield  {author} {\bibinfo {author} {\bibfnamefont {G.}~\bibnamefont
  {Grissonnanche}}, \bibinfo {author} {\bibfnamefont {O.}~\bibnamefont
  {Cyr-Choini{\`e}re}}, \bibinfo {author} {\bibfnamefont {F.}~\bibnamefont
  {Lalibert{\'e}}}, \bibinfo {author} {\bibfnamefont {S.}~\bibnamefont
  {Ren{\'e}de~Cotret}}, \bibinfo {author} {\bibfnamefont {A.}~\bibnamefont
  {Juneau-Fecteau}}, \bibinfo {author} {\bibfnamefont {S.}~\bibnamefont
  {Dufour-Beaus{\'e}jour}}, \bibinfo {author} {\bibfnamefont {M.~{\`E}.}\
  \bibnamefont {Delage}}, \bibinfo {author} {\bibfnamefont {D.}~\bibnamefont
  {LeBoeuf}}, \bibinfo {author} {\bibfnamefont {J.}~\bibnamefont {Chang}},
  \bibinfo {author} {\bibfnamefont {B.~J.}\ \bibnamefont {Ramshaw}}, \bibinfo
  {author} {\bibfnamefont {D.~A.}\ \bibnamefont {Bonn}}, \bibinfo {author}
  {\bibfnamefont {W.~N.}\ \bibnamefont {Hardy}}, \bibinfo {author}
  {\bibfnamefont {R.}~\bibnamefont {Liang}}, \bibinfo {author} {\bibfnamefont
  {S.}~\bibnamefont {Adachi}}, \bibinfo {author} {\bibfnamefont {N.~E.}\
  \bibnamefont {Hussey}}, \bibinfo {author} {\bibfnamefont {B.}~\bibnamefont
  {Vignolle}}, \bibinfo {author} {\bibfnamefont {C.}~\bibnamefont {Proust}},
  \bibinfo {author} {\bibfnamefont {M.}~\bibnamefont {Sutherland}}, \bibinfo
  {author} {\bibfnamefont {S.}~\bibnamefont {Kr{\"a}mer}}, \bibinfo {author}
  {\bibfnamefont {J.~H.}\ \bibnamefont {Park}}, \bibinfo {author}
  {\bibfnamefont {D.}~\bibnamefont {Graf}}, \bibinfo {author} {\bibfnamefont
  {N.}~\bibnamefont {Doiron-Leyraud}}, \ and\ \bibinfo {author} {\bibfnamefont
  {L.}~\bibnamefont {Taillefer}},\ }\href
  {http://dx.doi.org/10.1038/ncomms4280} {\bibfield  {journal} {\bibinfo
  {journal} {Nature Communications}\ }\textbf {\bibinfo {volume} {5}},\
  \bibinfo {pages} {3280 EP } (\bibinfo {year} {2014})}\BibitemShut {NoStop}%
\bibitem [{\citenamefont {Nie}\ \emph {et~al.}(2015)\citenamefont {Nie},
  \citenamefont {Sierens}, \citenamefont {Melko}, \citenamefont {Sachdev},\
  and\ \citenamefont {Kivelson}}]{Nie2015}%
  \BibitemOpen
  \bibfield  {author} {\bibinfo {author} {\bibfnamefont {L.}~\bibnamefont
  {Nie}}, \bibinfo {author} {\bibfnamefont {L.~E.~H.}\ \bibnamefont {Sierens}},
  \bibinfo {author} {\bibfnamefont {R.~G.}\ \bibnamefont {Melko}}, \bibinfo
  {author} {\bibfnamefont {S.}~\bibnamefont {Sachdev}}, \ and\ \bibinfo
  {author} {\bibfnamefont {S.~A.}\ \bibnamefont {Kivelson}},\ }\href {\doibase
  10.1103/PhysRevB.92.174505} {\bibfield  {journal} {\bibinfo  {journal} {Phys.
  Rev. B}\ }\textbf {\bibinfo {volume} {92}},\ \bibinfo {pages} {174505}
  (\bibinfo {year} {2015})}\BibitemShut {NoStop}%
\bibitem [{\citenamefont {Machida}\ \emph {et~al.}(2016)\citenamefont
  {Machida}, \citenamefont {Kohsaka}, \citenamefont {Matsuoka}, \citenamefont
  {Iwaya}, \citenamefont {Hanaguri},\ and\ \citenamefont
  {Tamegai}}]{Machida2015}%
  \BibitemOpen
  \bibfield  {author} {\bibinfo {author} {\bibfnamefont {T.}~\bibnamefont
  {Machida}}, \bibinfo {author} {\bibfnamefont {Y.}~\bibnamefont {Kohsaka}},
  \bibinfo {author} {\bibfnamefont {K.}~\bibnamefont {Matsuoka}}, \bibinfo
  {author} {\bibfnamefont {K.}~\bibnamefont {Iwaya}}, \bibinfo {author}
  {\bibfnamefont {T.}~\bibnamefont {Hanaguri}}, \ and\ \bibinfo {author}
  {\bibfnamefont {T.}~\bibnamefont {Tamegai}},\ }\href
  {http://dx.doi.org/10.1038/ncomms11747} {\bibfield  {journal} {\bibinfo
  {journal} {Nature Communications}\ }\textbf {\bibinfo {volume} {7}},\
  \bibinfo {pages} {11747 EP } (\bibinfo {year} {2016})}\BibitemShut {NoStop}%
\bibitem [{\citenamefont {Gerber}\ \emph {et~al.}(2015)\citenamefont {Gerber},
  \citenamefont {Jang}, \citenamefont {Nojiri}, \citenamefont {Matsuzawa},
  \citenamefont {Yasumura}, \citenamefont {Bonn}, \citenamefont {Liang},
  \citenamefont {Hardy}, \citenamefont {Islam}, \citenamefont {Mehta},
  \citenamefont {Song}, \citenamefont {Sikorski}, \citenamefont {Stefanescu},
  \citenamefont {Feng}, \citenamefont {Kivelson}, \citenamefont {Devereaux},
  \citenamefont {Shen}, \citenamefont {Kao}, \citenamefont {Lee}, \citenamefont
  {Zhu},\ and\ \citenamefont {Lee}}]{Gerber2015}%
  \BibitemOpen
  \bibfield  {author} {\bibinfo {author} {\bibfnamefont {S.}~\bibnamefont
  {Gerber}}, \bibinfo {author} {\bibfnamefont {H.}~\bibnamefont {Jang}},
  \bibinfo {author} {\bibfnamefont {H.}~\bibnamefont {Nojiri}}, \bibinfo
  {author} {\bibfnamefont {S.}~\bibnamefont {Matsuzawa}}, \bibinfo {author}
  {\bibfnamefont {H.}~\bibnamefont {Yasumura}}, \bibinfo {author}
  {\bibfnamefont {D.~A.}\ \bibnamefont {Bonn}}, \bibinfo {author}
  {\bibfnamefont {R.}~\bibnamefont {Liang}}, \bibinfo {author} {\bibfnamefont
  {W.~N.}\ \bibnamefont {Hardy}}, \bibinfo {author} {\bibfnamefont
  {Z.}~\bibnamefont {Islam}}, \bibinfo {author} {\bibfnamefont
  {A.}~\bibnamefont {Mehta}}, \bibinfo {author} {\bibfnamefont
  {S.}~\bibnamefont {Song}}, \bibinfo {author} {\bibfnamefont {M.}~\bibnamefont
  {Sikorski}}, \bibinfo {author} {\bibfnamefont {D.}~\bibnamefont
  {Stefanescu}}, \bibinfo {author} {\bibfnamefont {Y.}~\bibnamefont {Feng}},
  \bibinfo {author} {\bibfnamefont {S.~A.}\ \bibnamefont {Kivelson}}, \bibinfo
  {author} {\bibfnamefont {T.~P.}\ \bibnamefont {Devereaux}}, \bibinfo {author}
  {\bibfnamefont {Z.-X.}\ \bibnamefont {Shen}}, \bibinfo {author}
  {\bibfnamefont {C.-C.}\ \bibnamefont {Kao}}, \bibinfo {author} {\bibfnamefont
  {W.-S.}\ \bibnamefont {Lee}}, \bibinfo {author} {\bibfnamefont
  {D.}~\bibnamefont {Zhu}}, \ and\ \bibinfo {author} {\bibfnamefont {J.-S.}\
  \bibnamefont {Lee}},\ }\href {\doibase 10.1126/science.aac6257} {\bibfield
  {journal} {\bibinfo  {journal} {Science}\ }\textbf {\bibinfo {volume}
  {350}},\ \bibinfo {pages} {949} (\bibinfo {year} {2015})}\BibitemShut
  {NoStop}%
\bibitem [{\citenamefont {Chang}\ \emph {et~al.}(2016)\citenamefont {Chang},
  \citenamefont {Blackburn}, \citenamefont {Ivashko}, \citenamefont {Holmes},
  \citenamefont {Christensen}, \citenamefont {H{\"u}cker}, \citenamefont
  {Liang}, \citenamefont {Bonn}, \citenamefont {Hardy}, \citenamefont
  {R{\"u}tt}, \citenamefont {Zimmermann}, \citenamefont {Forgan},\ and\
  \citenamefont {Hayden}}]{Chang2016}%
  \BibitemOpen
  \bibfield  {author} {\bibinfo {author} {\bibfnamefont {J.}~\bibnamefont
  {Chang}}, \bibinfo {author} {\bibfnamefont {E.}~\bibnamefont {Blackburn}},
  \bibinfo {author} {\bibfnamefont {O.}~\bibnamefont {Ivashko}}, \bibinfo
  {author} {\bibfnamefont {A.~T.}\ \bibnamefont {Holmes}}, \bibinfo {author}
  {\bibfnamefont {N.~B.}\ \bibnamefont {Christensen}}, \bibinfo {author}
  {\bibfnamefont {M.}~\bibnamefont {H{\"u}cker}}, \bibinfo {author}
  {\bibfnamefont {R.}~\bibnamefont {Liang}}, \bibinfo {author} {\bibfnamefont
  {D.~A.}\ \bibnamefont {Bonn}}, \bibinfo {author} {\bibfnamefont {W.~N.}\
  \bibnamefont {Hardy}}, \bibinfo {author} {\bibfnamefont {U.}~\bibnamefont
  {R{\"u}tt}}, \bibinfo {author} {\bibfnamefont {M.~v.}\ \bibnamefont
  {Zimmermann}}, \bibinfo {author} {\bibfnamefont {E.~M.}\ \bibnamefont
  {Forgan}}, \ and\ \bibinfo {author} {\bibfnamefont {S.~M.}\ \bibnamefont
  {Hayden}},\ }\href {https://doi.org/10.1038/ncomms11494} {\bibfield
  {journal} {\bibinfo  {journal} {Nature Communications}\ }\textbf {\bibinfo
  {volume} {7}},\ \bibinfo {pages} {11494 EP } (\bibinfo {year}
  {2016})}\BibitemShut {NoStop}%
\bibitem [{\citenamefont {Yu}\ \emph {et~al.}(2016)\citenamefont {Yu},
  \citenamefont {Hirschberger}, \citenamefont {Loew}, \citenamefont {Li},
  \citenamefont {Lawson}, \citenamefont {Asaba}, \citenamefont {Kemper},
  \citenamefont {Liang}, \citenamefont {Porras}, \citenamefont {Boebinger},
  \citenamefont {Singleton}, \citenamefont {Keimer}, \citenamefont {Li},\ and\
  \citenamefont {Ong}}]{Yu2016}%
  \BibitemOpen
  \bibfield  {author} {\bibinfo {author} {\bibfnamefont {F.}~\bibnamefont
  {Yu}}, \bibinfo {author} {\bibfnamefont {M.}~\bibnamefont {Hirschberger}},
  \bibinfo {author} {\bibfnamefont {T.}~\bibnamefont {Loew}}, \bibinfo {author}
  {\bibfnamefont {G.}~\bibnamefont {Li}}, \bibinfo {author} {\bibfnamefont
  {B.~J.}\ \bibnamefont {Lawson}}, \bibinfo {author} {\bibfnamefont
  {T.}~\bibnamefont {Asaba}}, \bibinfo {author} {\bibfnamefont {J.~B.}\
  \bibnamefont {Kemper}}, \bibinfo {author} {\bibfnamefont {T.}~\bibnamefont
  {Liang}}, \bibinfo {author} {\bibfnamefont {J.}~\bibnamefont {Porras}},
  \bibinfo {author} {\bibfnamefont {G.~S.}\ \bibnamefont {Boebinger}}, \bibinfo
  {author} {\bibfnamefont {J.}~\bibnamefont {Singleton}}, \bibinfo {author}
  {\bibfnamefont {B.}~\bibnamefont {Keimer}}, \bibinfo {author} {\bibfnamefont
  {L.}~\bibnamefont {Li}}, \ and\ \bibinfo {author} {\bibfnamefont {N.~P.}\
  \bibnamefont {Ong}},\ }\href {\doibase 10.1073/pnas.1612591113} {\bibfield
  {journal} {\bibinfo  {journal} {Proceedings of the National Academy of
  Sciences}\ }\textbf {\bibinfo {volume} {113}},\ \bibinfo {pages} {12667}
  (\bibinfo {year} {2016})}\BibitemShut {NoStop}%
\bibitem [{\citenamefont {Jang}\ \emph {et~al.}(2016)\citenamefont {Jang},
  \citenamefont {Lee}, \citenamefont {Nojiri}, \citenamefont {Matsuzawa},
  \citenamefont {Yasumura}, \citenamefont {Nie}, \citenamefont {Maharaj},
  \citenamefont {Gerber}, \citenamefont {Liu}, \citenamefont {Mehta},
  \citenamefont {Bonn}, \citenamefont {Liang}, \citenamefont {Hardy},
  \citenamefont {Burns}, \citenamefont {Islam}, \citenamefont {Song},
  \citenamefont {Hastings}, \citenamefont {Devereaux}, \citenamefont {Shen},
  \citenamefont {Kivelson}, \citenamefont {Kao}, \citenamefont {Zhu},\ and\
  \citenamefont {Lee}}]{Jang2016}%
  \BibitemOpen
  \bibfield  {author} {\bibinfo {author} {\bibfnamefont {H.}~\bibnamefont
  {Jang}}, \bibinfo {author} {\bibfnamefont {W.-S.}\ \bibnamefont {Lee}},
  \bibinfo {author} {\bibfnamefont {H.}~\bibnamefont {Nojiri}}, \bibinfo
  {author} {\bibfnamefont {S.}~\bibnamefont {Matsuzawa}}, \bibinfo {author}
  {\bibfnamefont {H.}~\bibnamefont {Yasumura}}, \bibinfo {author}
  {\bibfnamefont {L.}~\bibnamefont {Nie}}, \bibinfo {author} {\bibfnamefont
  {A.~V.}\ \bibnamefont {Maharaj}}, \bibinfo {author} {\bibfnamefont
  {S.}~\bibnamefont {Gerber}}, \bibinfo {author} {\bibfnamefont {Y.-J.}\
  \bibnamefont {Liu}}, \bibinfo {author} {\bibfnamefont {A.}~\bibnamefont
  {Mehta}}, \bibinfo {author} {\bibfnamefont {D.~A.}\ \bibnamefont {Bonn}},
  \bibinfo {author} {\bibfnamefont {R.}~\bibnamefont {Liang}}, \bibinfo
  {author} {\bibfnamefont {W.~N.}\ \bibnamefont {Hardy}}, \bibinfo {author}
  {\bibfnamefont {C.~A.}\ \bibnamefont {Burns}}, \bibinfo {author}
  {\bibfnamefont {Z.}~\bibnamefont {Islam}}, \bibinfo {author} {\bibfnamefont
  {S.}~\bibnamefont {Song}}, \bibinfo {author} {\bibfnamefont {J.}~\bibnamefont
  {Hastings}}, \bibinfo {author} {\bibfnamefont {T.~P.}\ \bibnamefont
  {Devereaux}}, \bibinfo {author} {\bibfnamefont {Z.-X.}\ \bibnamefont {Shen}},
  \bibinfo {author} {\bibfnamefont {S.~A.}\ \bibnamefont {Kivelson}}, \bibinfo
  {author} {\bibfnamefont {C.-C.}\ \bibnamefont {Kao}}, \bibinfo {author}
  {\bibfnamefont {D.}~\bibnamefont {Zhu}}, \ and\ \bibinfo {author}
  {\bibfnamefont {J.-S.}\ \bibnamefont {Lee}},\ }\href {\doibase
  10.1073/pnas.1612849113} {\bibfield  {journal} {\bibinfo  {journal}
  {Proceedings of the National Academy of Sciences}\ }\textbf {\bibinfo
  {volume} {113}},\ \bibinfo {pages} {14645} (\bibinfo {year}
  {2016})}\BibitemShut {NoStop}%
\bibitem [{\citenamefont {Leroux}\ \emph {et~al.}(2019)\citenamefont {Leroux},
  \citenamefont {Mishra}, \citenamefont {Ruff}, \citenamefont {Claus},
  \citenamefont {Smylie}, \citenamefont {Opagiste}, \citenamefont
  {Rodi{\`e}re}, \citenamefont {Kayani}, \citenamefont {Gu}, \citenamefont
  {Tranquada}, \citenamefont {Kwok}, \citenamefont {Islam},\ and\ \citenamefont
  {Welp}}]{Leroux2019}%
  \BibitemOpen
  \bibfield  {author} {\bibinfo {author} {\bibfnamefont {M.}~\bibnamefont
  {Leroux}}, \bibinfo {author} {\bibfnamefont {V.}~\bibnamefont {Mishra}},
  \bibinfo {author} {\bibfnamefont {J.~P.~C.}\ \bibnamefont {Ruff}}, \bibinfo
  {author} {\bibfnamefont {H.}~\bibnamefont {Claus}}, \bibinfo {author}
  {\bibfnamefont {M.~P.}\ \bibnamefont {Smylie}}, \bibinfo {author}
  {\bibfnamefont {C.}~\bibnamefont {Opagiste}}, \bibinfo {author}
  {\bibfnamefont {P.}~\bibnamefont {Rodi{\`e}re}}, \bibinfo {author}
  {\bibfnamefont {A.}~\bibnamefont {Kayani}}, \bibinfo {author} {\bibfnamefont
  {G.~D.}\ \bibnamefont {Gu}}, \bibinfo {author} {\bibfnamefont {J.~M.}\
  \bibnamefont {Tranquada}}, \bibinfo {author} {\bibfnamefont {W.-K.}\
  \bibnamefont {Kwok}}, \bibinfo {author} {\bibfnamefont {Z.}~\bibnamefont
  {Islam}}, \ and\ \bibinfo {author} {\bibfnamefont {U.}~\bibnamefont {Welp}},\
  }\href {\doibase 10.1073/pnas.1817134116} {\bibfield  {journal} {\bibinfo
  {journal} {Proceedings of the National Academy of Sciences}\ }\textbf
  {\bibinfo {volume} {116}},\ \bibinfo {pages} {10691} (\bibinfo {year}
  {2019})},\ \Eprint
  {http://arxiv.org/abs/https://www.pnas.org/content/116/22/10691.full.pdf}
  {https://www.pnas.org/content/116/22/10691.full.pdf} \BibitemShut {NoStop}%
\bibitem [{\citenamefont {Morosan}\ \emph {et~al.}(2006)\citenamefont
  {Morosan}, \citenamefont {Zandbergen}, \citenamefont {Dennis}, \citenamefont
  {Bos}, \citenamefont {Onose}, \citenamefont {Klimczuk}, \citenamefont
  {Ramirez}, \citenamefont {Ong},\ and\ \citenamefont {Cava}}]{Morosan2006}%
  \BibitemOpen
  \bibfield  {author} {\bibinfo {author} {\bibfnamefont {E.}~\bibnamefont
  {Morosan}}, \bibinfo {author} {\bibfnamefont {H.~W.}\ \bibnamefont
  {Zandbergen}}, \bibinfo {author} {\bibfnamefont {B.~S.}\ \bibnamefont
  {Dennis}}, \bibinfo {author} {\bibfnamefont {J.~W.~G.}\ \bibnamefont {Bos}},
  \bibinfo {author} {\bibfnamefont {Y.}~\bibnamefont {Onose}}, \bibinfo
  {author} {\bibfnamefont {T.}~\bibnamefont {Klimczuk}}, \bibinfo {author}
  {\bibfnamefont {A.~P.}\ \bibnamefont {Ramirez}}, \bibinfo {author}
  {\bibfnamefont {N.~P.}\ \bibnamefont {Ong}}, \ and\ \bibinfo {author}
  {\bibfnamefont {R.~J.}\ \bibnamefont {Cava}},\ }\href {\doibase
  10.1038/nphys360} {\bibfield  {journal} {\bibinfo  {journal} {Nature
  Physics}\ }\textbf {\bibinfo {volume} {2}},\ \bibinfo {pages} {544} (\bibinfo
  {year} {2006})}\BibitemShut {NoStop}%
\bibitem [{\citenamefont {Kusmartseva}\ \emph {et~al.}(2009)\citenamefont
  {Kusmartseva}, \citenamefont {Sipos}, \citenamefont {Berger}, \citenamefont
  {Forr\'o},\ and\ \citenamefont {Tuti\ifmmode~\check{s}\else
  \v{s}\fi{}}}]{Kusmartseva2009}%
  \BibitemOpen
  \bibfield  {author} {\bibinfo {author} {\bibfnamefont {A.~F.}\ \bibnamefont
  {Kusmartseva}}, \bibinfo {author} {\bibfnamefont {B.}~\bibnamefont {Sipos}},
  \bibinfo {author} {\bibfnamefont {H.}~\bibnamefont {Berger}}, \bibinfo
  {author} {\bibfnamefont {L.}~\bibnamefont {Forr\'o}}, \ and\ \bibinfo
  {author} {\bibfnamefont {E.}~\bibnamefont {Tuti\ifmmode~\check{s}\else
  \v{s}\fi{}}},\ }\href {\doibase 10.1103/PhysRevLett.103.236401} {\bibfield
  {journal} {\bibinfo  {journal} {Phys. Rev. Lett.}\ }\textbf {\bibinfo
  {volume} {103}},\ \bibinfo {pages} {236401} (\bibinfo {year}
  {2009})}\BibitemShut {NoStop}%
\bibitem [{\citenamefont {Kiss}\ \emph {et~al.}(2007)\citenamefont {Kiss},
  \citenamefont {Yokoya}, \citenamefont {Chainani}, \citenamefont {Shin},
  \citenamefont {Hanaguri}, \citenamefont {Nohara},\ and\ \citenamefont
  {Takagi}}]{Kiss2007}%
  \BibitemOpen
  \bibfield  {author} {\bibinfo {author} {\bibfnamefont {T.}~\bibnamefont
  {Kiss}}, \bibinfo {author} {\bibfnamefont {T.}~\bibnamefont {Yokoya}},
  \bibinfo {author} {\bibfnamefont {A.}~\bibnamefont {Chainani}}, \bibinfo
  {author} {\bibfnamefont {S.}~\bibnamefont {Shin}}, \bibinfo {author}
  {\bibfnamefont {T.}~\bibnamefont {Hanaguri}}, \bibinfo {author}
  {\bibfnamefont {M.}~\bibnamefont {Nohara}}, \ and\ \bibinfo {author}
  {\bibfnamefont {H.}~\bibnamefont {Takagi}},\ }\href {\doibase
  10.1038/nphys699} {\bibfield  {journal} {\bibinfo  {journal} {Nature
  Physics}\ }\textbf {\bibinfo {volume} {3}},\ \bibinfo {pages} {720} (\bibinfo
  {year} {2007})}\BibitemShut {NoStop}%
\bibitem [{\citenamefont {Liu}\ \emph {et~al.}(2016)\citenamefont {Liu},
  \citenamefont {Shao}, \citenamefont {Li}, \citenamefont {Lu}, \citenamefont
  {Zhu}, \citenamefont {Tong}, \citenamefont {Xiao}, \citenamefont {Ling},
  \citenamefont {Xi}, \citenamefont {Pi}, \citenamefont {Tian}, \citenamefont
  {Yang}, \citenamefont {Li}, \citenamefont {Song}, \citenamefont {Zhu},\ and\
  \citenamefont {Sun}}]{Liu2016}%
  \BibitemOpen
  \bibfield  {author} {\bibinfo {author} {\bibfnamefont {Y.}~\bibnamefont
  {Liu}}, \bibinfo {author} {\bibfnamefont {D.~F.}\ \bibnamefont {Shao}},
  \bibinfo {author} {\bibfnamefont {L.~J.}\ \bibnamefont {Li}}, \bibinfo
  {author} {\bibfnamefont {W.~J.}\ \bibnamefont {Lu}}, \bibinfo {author}
  {\bibfnamefont {X.~D.}\ \bibnamefont {Zhu}}, \bibinfo {author} {\bibfnamefont
  {P.}~\bibnamefont {Tong}}, \bibinfo {author} {\bibfnamefont {R.~C.}\
  \bibnamefont {Xiao}}, \bibinfo {author} {\bibfnamefont {L.~S.}\ \bibnamefont
  {Ling}}, \bibinfo {author} {\bibfnamefont {C.~Y.}\ \bibnamefont {Xi}},
  \bibinfo {author} {\bibfnamefont {L.}~\bibnamefont {Pi}}, \bibinfo {author}
  {\bibfnamefont {H.~F.}\ \bibnamefont {Tian}}, \bibinfo {author}
  {\bibfnamefont {H.~X.}\ \bibnamefont {Yang}}, \bibinfo {author}
  {\bibfnamefont {J.~Q.}\ \bibnamefont {Li}}, \bibinfo {author} {\bibfnamefont
  {W.~H.}\ \bibnamefont {Song}}, \bibinfo {author} {\bibfnamefont {X.~B.}\
  \bibnamefont {Zhu}}, \ and\ \bibinfo {author} {\bibfnamefont {Y.~P.}\
  \bibnamefont {Sun}},\ }\href {\doibase 10.1103/PhysRevB.94.045131} {\bibfield
   {journal} {\bibinfo  {journal} {Phys. Rev. B}\ }\textbf {\bibinfo {volume}
  {94}},\ \bibinfo {pages} {045131} (\bibinfo {year} {2016})}\BibitemShut
  {NoStop}%
\bibitem [{\citenamefont {Cho}\ \emph {et~al.}(2018)\citenamefont {Cho},
  \citenamefont {Ko{\'n}czykowski}, \citenamefont {Teknowijoyo}, \citenamefont
  {Tanatar}, \citenamefont {Guss}, \citenamefont {Gartin}, \citenamefont
  {Wilde}, \citenamefont {Kreyssig}, \citenamefont {McQueeney}, \citenamefont
  {Goldman}, \citenamefont {Mishra}, \citenamefont {Hirschfeld},\ and\
  \citenamefont {Prozorov}}]{Cho2018}%
  \BibitemOpen
  \bibfield  {author} {\bibinfo {author} {\bibfnamefont {K.}~\bibnamefont
  {Cho}}, \bibinfo {author} {\bibfnamefont {M.}~\bibnamefont
  {Ko{\'n}czykowski}}, \bibinfo {author} {\bibfnamefont {S.}~\bibnamefont
  {Teknowijoyo}}, \bibinfo {author} {\bibfnamefont {M.~A.}\ \bibnamefont
  {Tanatar}}, \bibinfo {author} {\bibfnamefont {J.}~\bibnamefont {Guss}},
  \bibinfo {author} {\bibfnamefont {P.~B.}\ \bibnamefont {Gartin}}, \bibinfo
  {author} {\bibfnamefont {J.~M.}\ \bibnamefont {Wilde}}, \bibinfo {author}
  {\bibfnamefont {A.}~\bibnamefont {Kreyssig}}, \bibinfo {author}
  {\bibfnamefont {R.~J.}\ \bibnamefont {McQueeney}}, \bibinfo {author}
  {\bibfnamefont {A.~I.}\ \bibnamefont {Goldman}}, \bibinfo {author}
  {\bibfnamefont {V.}~\bibnamefont {Mishra}}, \bibinfo {author} {\bibfnamefont
  {P.~J.}\ \bibnamefont {Hirschfeld}}, \ and\ \bibinfo {author} {\bibfnamefont
  {R.}~\bibnamefont {Prozorov}},\ }\href {\doibase 10.1038/s41467-018-05153-0}
  {\bibfield  {journal} {\bibinfo  {journal} {Nature Communications}\ }\textbf
  {\bibinfo {volume} {9}},\ \bibinfo {pages} {2796} (\bibinfo {year}
  {2018})}\BibitemShut {NoStop}%
\bibitem [{\citenamefont {Yang}\ \emph {et~al.}(2018)\citenamefont {Yang},
  \citenamefont {Fang}, \citenamefont {Fatemi}, \citenamefont {Ruhman},
  \citenamefont {Navarro-Moratalla}, \citenamefont {Watanabe}, \citenamefont
  {Taniguchi}, \citenamefont {Kaxiras},\ and\ \citenamefont
  {Jarillo-Herrero}}]{Yang2018}%
  \BibitemOpen
  \bibfield  {author} {\bibinfo {author} {\bibfnamefont {Y.}~\bibnamefont
  {Yang}}, \bibinfo {author} {\bibfnamefont {S.}~\bibnamefont {Fang}}, \bibinfo
  {author} {\bibfnamefont {V.}~\bibnamefont {Fatemi}}, \bibinfo {author}
  {\bibfnamefont {J.}~\bibnamefont {Ruhman}}, \bibinfo {author} {\bibfnamefont
  {E.}~\bibnamefont {Navarro-Moratalla}}, \bibinfo {author} {\bibfnamefont
  {K.}~\bibnamefont {Watanabe}}, \bibinfo {author} {\bibfnamefont
  {T.}~\bibnamefont {Taniguchi}}, \bibinfo {author} {\bibfnamefont
  {E.}~\bibnamefont {Kaxiras}}, \ and\ \bibinfo {author} {\bibfnamefont
  {P.}~\bibnamefont {Jarillo-Herrero}},\ }\href {\doibase
  10.1103/PhysRevB.98.035203} {\bibfield  {journal} {\bibinfo  {journal} {Phys.
  Rev. B}\ }\textbf {\bibinfo {volume} {98}},\ \bibinfo {pages} {035203}
  (\bibinfo {year} {2018})}\BibitemShut {NoStop}%
\bibitem [{\citenamefont {Lee}\ \emph {et~al.}(2019)\citenamefont {Lee},
  \citenamefont {de~la Pe\~na}, \citenamefont {Sun}, \citenamefont {Mitrano},
  \citenamefont {Fang}, \citenamefont {Jang}, \citenamefont {Lee},
  \citenamefont {Eckberg}, \citenamefont {Campbell}, \citenamefont {Collini},
  \citenamefont {Paglione}, \citenamefont {de~Groot},\ and\ \citenamefont
  {Abbamonte}}]{Lee2019}%
  \BibitemOpen
  \bibfield  {author} {\bibinfo {author} {\bibfnamefont {S.}~\bibnamefont
  {Lee}}, \bibinfo {author} {\bibfnamefont {G.}~\bibnamefont {de~la Pe\~na}},
  \bibinfo {author} {\bibfnamefont {S.~X.-L.}\ \bibnamefont {Sun}}, \bibinfo
  {author} {\bibfnamefont {M.}~\bibnamefont {Mitrano}}, \bibinfo {author}
  {\bibfnamefont {Y.}~\bibnamefont {Fang}}, \bibinfo {author} {\bibfnamefont
  {H.}~\bibnamefont {Jang}}, \bibinfo {author} {\bibfnamefont {J.-S.}\
  \bibnamefont {Lee}}, \bibinfo {author} {\bibfnamefont {C.}~\bibnamefont
  {Eckberg}}, \bibinfo {author} {\bibfnamefont {D.}~\bibnamefont {Campbell}},
  \bibinfo {author} {\bibfnamefont {J.}~\bibnamefont {Collini}}, \bibinfo
  {author} {\bibfnamefont {J.}~\bibnamefont {Paglione}}, \bibinfo {author}
  {\bibfnamefont {F.~M.~F.}\ \bibnamefont {de~Groot}}, \ and\ \bibinfo {author}
  {\bibfnamefont {P.}~\bibnamefont {Abbamonte}},\ }\href {\doibase
  10.1103/PhysRevLett.122.147601} {\bibfield  {journal} {\bibinfo  {journal}
  {Phys. Rev. Lett.}\ }\textbf {\bibinfo {volume} {122}},\ \bibinfo {pages}
  {147601} (\bibinfo {year} {2019})}\BibitemShut {NoStop}%
\bibitem [{\citenamefont {Sleight}\ \emph {et~al.}(1975)\citenamefont
  {Sleight}, \citenamefont {Gillson},\ and\ \citenamefont
  {Bierstedt}}]{Sleight1975}%
  \BibitemOpen
  \bibfield  {author} {\bibinfo {author} {\bibfnamefont {A.}~\bibnamefont
  {Sleight}}, \bibinfo {author} {\bibfnamefont {J.}~\bibnamefont {Gillson}}, \
  and\ \bibinfo {author} {\bibfnamefont {P.}~\bibnamefont {Bierstedt}},\ }\href
  {\doibase https://doi.org/10.1016/0038-1098(75)90327-0} {\bibfield  {journal}
  {\bibinfo  {journal} {Solid State Communications}\ }\textbf {\bibinfo
  {volume} {17}},\ \bibinfo {pages} {27 } (\bibinfo {year} {1975})}\BibitemShut
  {NoStop}%
\bibitem [{\citenamefont {Cava}\ \emph {et~al.}(1988)\citenamefont {Cava},
  \citenamefont {Batlogg}, \citenamefont {Krajewski}, \citenamefont {Farrow},
  \citenamefont {Rupp}, \citenamefont {White}, \citenamefont {Short},
  \citenamefont {Peck},\ and\ \citenamefont {Kometani}}]{Cava1988}%
  \BibitemOpen
  \bibfield  {author} {\bibinfo {author} {\bibfnamefont {R.~J.}\ \bibnamefont
  {Cava}}, \bibinfo {author} {\bibfnamefont {B.}~\bibnamefont {Batlogg}},
  \bibinfo {author} {\bibfnamefont {J.~J.}\ \bibnamefont {Krajewski}}, \bibinfo
  {author} {\bibfnamefont {R.}~\bibnamefont {Farrow}}, \bibinfo {author}
  {\bibfnamefont {L.~W.}\ \bibnamefont {Rupp}}, \bibinfo {author}
  {\bibfnamefont {A.~E.}\ \bibnamefont {White}}, \bibinfo {author}
  {\bibfnamefont {K.}~\bibnamefont {Short}}, \bibinfo {author} {\bibfnamefont
  {W.~F.}\ \bibnamefont {Peck}}, \ and\ \bibinfo {author} {\bibfnamefont
  {T.}~\bibnamefont {Kometani}},\ }\href {\doibase 10.1038/332814a0} {\bibfield
   {journal} {\bibinfo  {journal} {Nature}\ }\textbf {\bibinfo {volume}
  {332}},\ \bibinfo {pages} {814} (\bibinfo {year} {1988})}\BibitemShut
  {NoStop}%
\bibitem [{\citenamefont {Sleight}(2015)}]{Sleight2015}%
  \BibitemOpen
  \bibfield  {author} {\bibinfo {author} {\bibfnamefont {A.~W.}\ \bibnamefont
  {Sleight}},\ }\href {\doibase https://doi.org/10.1016/j.physc.2015.02.012}
  {\bibfield  {journal} {\bibinfo  {journal} {Physica C: Superconductivity and
  its Applications}\ }\textbf {\bibinfo {volume} {514}},\ \bibinfo {pages} {152
  } (\bibinfo {year} {2015})},\ \bibinfo {note} {superconducting Materials:
  Conventional, Unconventional and Undetermined}\BibitemShut {NoStop}%
\bibitem [{\citenamefont {Karmakar}\ \emph {et~al.}(2017)\citenamefont
  {Karmakar}, \citenamefont {Menon},\ and\ \citenamefont
  {Ganesh}}]{KarmakarPRB2017}%
  \BibitemOpen
  \bibfield  {author} {\bibinfo {author} {\bibfnamefont {M.}~\bibnamefont
  {Karmakar}}, \bibinfo {author} {\bibfnamefont {G.~I.}\ \bibnamefont {Menon}},
  \ and\ \bibinfo {author} {\bibfnamefont {R.}~\bibnamefont {Ganesh}},\ }\href
  {\doibase 10.1103/PhysRevB.96.174501} {\bibfield  {journal} {\bibinfo
  {journal} {Phys. Rev. B}\ }\textbf {\bibinfo {volume} {96}},\ \bibinfo
  {pages} {174501} (\bibinfo {year} {2017})}\BibitemShut {NoStop}%
\bibitem [{\citenamefont {Boninsegni}\ and\ \citenamefont
  {Prokof'ev}(2012)}]{Boninsegni2012}%
  \BibitemOpen
  \bibfield  {author} {\bibinfo {author} {\bibfnamefont {M.}~\bibnamefont
  {Boninsegni}}\ and\ \bibinfo {author} {\bibfnamefont {N.~V.}\ \bibnamefont
  {Prokof'ev}},\ }\href {\doibase 10.1103/RevModPhys.84.759} {\bibfield
  {journal} {\bibinfo  {journal} {Rev. Mod. Phys.}\ }\textbf {\bibinfo {volume}
  {84}},\ \bibinfo {pages} {759} (\bibinfo {year} {2012})}\BibitemShut
  {NoStop}%
\bibitem [{\citenamefont {Karmakar}\ and\ \citenamefont
  {Ganesh}(2017)}]{KarmakarJPSJ2017}%
  \BibitemOpen
  \bibfield  {author} {\bibinfo {author} {\bibfnamefont {M.}~\bibnamefont
  {Karmakar}}\ and\ \bibinfo {author} {\bibfnamefont {R.}~\bibnamefont
  {Ganesh}},\ }\href {\doibase 10.7566/JPSJ.86.124719} {\bibfield  {journal}
  {\bibinfo  {journal} {Journal of the Physical Society of Japan}\ }\textbf
  {\bibinfo {volume} {86}},\ \bibinfo {pages} {124719} (\bibinfo {year}
  {2017})}\BibitemShut {NoStop}%
\bibitem [{\citenamefont {Saran}\ \emph {et~al.}(2019)\citenamefont {Saran},
  \citenamefont {Karmakar},\ and\ \citenamefont {Ganesh}}]{Saran2019}%
  \BibitemOpen
  \bibfield  {author} {\bibinfo {author} {\bibfnamefont {V.}~\bibnamefont
  {Saran}}, \bibinfo {author} {\bibfnamefont {M.}~\bibnamefont {Karmakar}}, \
  and\ \bibinfo {author} {\bibfnamefont {R.}~\bibnamefont {Ganesh}},\ }\href
  {\doibase 10.1103/PhysRevB.100.104520} {\bibfield  {journal} {\bibinfo
  {journal} {Phys. Rev. B}\ }\textbf {\bibinfo {volume} {100}},\ \bibinfo
  {pages} {104520} (\bibinfo {year} {2019})}\BibitemShut {NoStop}%
\bibitem [{\citenamefont {Demler}\ \emph {et~al.}(2004)\citenamefont {Demler},
  \citenamefont {Hanke},\ and\ \citenamefont {Zhang}}]{Demler2004}%
  \BibitemOpen
  \bibfield  {author} {\bibinfo {author} {\bibfnamefont {E.}~\bibnamefont
  {Demler}}, \bibinfo {author} {\bibfnamefont {W.}~\bibnamefont {Hanke}}, \
  and\ \bibinfo {author} {\bibfnamefont {S.-C.}\ \bibnamefont {Zhang}},\ }\href
  {\doibase 10.1103/RevModPhys.76.909} {\bibfield  {journal} {\bibinfo
  {journal} {Rev. Mod. Phys.}\ }\textbf {\bibinfo {volume} {76}},\ \bibinfo
  {pages} {909} (\bibinfo {year} {2004})}\BibitemShut {NoStop}%
\bibitem [{\citenamefont {Arovas}\ \emph {et~al.}(1997)\citenamefont {Arovas},
  \citenamefont {Berlinsky}, \citenamefont {Kallin},\ and\ \citenamefont
  {Zhang}}]{Arovas1997}%
  \BibitemOpen
  \bibfield  {author} {\bibinfo {author} {\bibfnamefont {D.~P.}\ \bibnamefont
  {Arovas}}, \bibinfo {author} {\bibfnamefont {A.~J.}\ \bibnamefont
  {Berlinsky}}, \bibinfo {author} {\bibfnamefont {C.}~\bibnamefont {Kallin}}, \
  and\ \bibinfo {author} {\bibfnamefont {S.-C.}\ \bibnamefont {Zhang}},\ }\href
  {\doibase 10.1103/PhysRevLett.79.2871} {\bibfield  {journal} {\bibinfo
  {journal} {Phys. Rev. Lett.}\ }\textbf {\bibinfo {volume} {79}},\ \bibinfo
  {pages} {2871} (\bibinfo {year} {1997})}\BibitemShut {NoStop}%
\bibitem [{\citenamefont {Hu}\ and\ \citenamefont {Zhang}(2002)}]{Hu2002}%
  \BibitemOpen
  \bibfield  {author} {\bibinfo {author} {\bibfnamefont {J.-P.}\ \bibnamefont
  {Hu}}\ and\ \bibinfo {author} {\bibfnamefont {S.-C.}\ \bibnamefont {Zhang}},\
  }\href {\doibase http://dx.doi.org/10.1016/S0022-3697(02)00243-3} {\bibfield
  {journal} {\bibinfo  {journal} {Journal of Physics and Chemistry of Solids}\
  }\textbf {\bibinfo {volume} {63}},\ \bibinfo {pages} {2277 } (\bibinfo {year}
  {2002})},\ \bibinfo {note} {proceedings of the Conference on Spectroscopies
  in Novel Superconductors}\BibitemShut {NoStop}%
\bibitem [{\citenamefont {Tarruell}\ and\ \citenamefont
  {Sanchez-Palencia}(2018)}]{Tarruell2018}%
  \BibitemOpen
  \bibfield  {author} {\bibinfo {author} {\bibfnamefont {L.}~\bibnamefont
  {Tarruell}}\ and\ \bibinfo {author} {\bibfnamefont {L.}~\bibnamefont
  {Sanchez-Palencia}},\ }\href {\doibase
  https://doi.org/10.1016/j.crhy.2018.10.013} {\bibfield  {journal} {\bibinfo
  {journal} {Comptes Rendus Physique}\ }\textbf {\bibinfo {volume} {19}},\
  \bibinfo {pages} {365 } (\bibinfo {year} {2018})},\ \bibinfo {note} {quantum
  simulation / Simulation quantique}\BibitemShut {NoStop}%
\bibitem [{\citenamefont {Yang}(1989)}]{Yang1989}%
  \BibitemOpen
  \bibfield  {author} {\bibinfo {author} {\bibfnamefont {C.~N.}\ \bibnamefont
  {Yang}},\ }\href {\doibase 10.1103/PhysRevLett.63.2144} {\bibfield  {journal}
  {\bibinfo  {journal} {Phys. Rev. Lett.}\ }\textbf {\bibinfo {volume} {63}},\
  \bibinfo {pages} {2144} (\bibinfo {year} {1989})}\BibitemShut {NoStop}%
\bibitem [{\citenamefont {Yang}\ and\ \citenamefont {Zhang}(1990)}]{Yang1990}%
  \BibitemOpen
  \bibfield  {author} {\bibinfo {author} {\bibfnamefont {C.~N.}\ \bibnamefont
  {Yang}}\ and\ \bibinfo {author} {\bibfnamefont {S.~C.}\ \bibnamefont
  {Zhang}},\ }\href {\doibase 10.1142/S0217984990000933} {\bibfield  {journal}
  {\bibinfo  {journal} {Modern Physics Letters B}\ }\textbf {\bibinfo {volume}
  {04}},\ \bibinfo {pages} {759} (\bibinfo {year} {1990})}\BibitemShut
  {NoStop}%
\bibitem [{\citenamefont {Zhang}(1990)}]{Zhang1990}%
  \BibitemOpen
  \bibfield  {author} {\bibinfo {author} {\bibfnamefont {S.}~\bibnamefont
  {Zhang}},\ }\href {\doibase 10.1103/PhysRevLett.65.120} {\bibfield  {journal}
  {\bibinfo  {journal} {Phys. Rev. Lett.}\ }\textbf {\bibinfo {volume} {65}},\
  \bibinfo {pages} {120} (\bibinfo {year} {1990})}\BibitemShut {NoStop}%
\bibitem [{\citenamefont {Ramachandran}(2011)}]{Ganeshthesis}%
  \BibitemOpen
  \bibfield  {author} {\bibinfo {author} {\bibfnamefont {G.}~\bibnamefont
  {Ramachandran}},\ }\emph {\bibinfo {title} {Competing Orders in Strongly
  Correlated Systems}},\ \href
  {https://tspace.library.utoronto.ca/handle/1807/32868} {Ph.D. thesis},\
  \bibinfo  {school} {University of Toronto} (\bibinfo {year} {2011}),\
  \bibinfo {note} {chapter 5}\BibitemShut {NoStop}%
\bibitem [{\citenamefont {Burkov}\ and\ \citenamefont
  {Paramekanti}(2008)}]{Burkov2008}%
  \BibitemOpen
  \bibfield  {author} {\bibinfo {author} {\bibfnamefont {A.~A.}\ \bibnamefont
  {Burkov}}\ and\ \bibinfo {author} {\bibfnamefont {A.}~\bibnamefont
  {Paramekanti}},\ }\href {\doibase 10.1103/PhysRevLett.100.255301} {\bibfield
  {journal} {\bibinfo  {journal} {Phys. Rev. Lett.}\ }\textbf {\bibinfo
  {volume} {100}},\ \bibinfo {pages} {255301} (\bibinfo {year}
  {2008})}\BibitemShut {NoStop}%
\bibitem [{\citenamefont {Ganesh}\ \emph {et~al.}(2009)\citenamefont {Ganesh},
  \citenamefont {Paramekanti},\ and\ \citenamefont {Burkov}}]{Ganesh2009}%
  \BibitemOpen
  \bibfield  {author} {\bibinfo {author} {\bibfnamefont {R.}~\bibnamefont
  {Ganesh}}, \bibinfo {author} {\bibfnamefont {A.}~\bibnamefont {Paramekanti}},
  \ and\ \bibinfo {author} {\bibfnamefont {A.~A.}\ \bibnamefont {Burkov}},\
  }\href {\doibase 10.1103/PhysRevA.80.043612} {\bibfield  {journal} {\bibinfo
  {journal} {Phys. Rev. A}\ }\textbf {\bibinfo {volume} {80}},\ \bibinfo
  {pages} {043612} (\bibinfo {year} {2009})}\BibitemShut {NoStop}%
\bibitem [{\citenamefont {Yunomae}\ \emph {et~al.}(2009)\citenamefont
  {Yunomae}, \citenamefont {Yamamoto}, \citenamefont {Danshita}, \citenamefont
  {Yokoshi},\ and\ \citenamefont {Tsuchiya}}]{Yunomae2009}%
  \BibitemOpen
  \bibfield  {author} {\bibinfo {author} {\bibfnamefont {Y.}~\bibnamefont
  {Yunomae}}, \bibinfo {author} {\bibfnamefont {D.}~\bibnamefont {Yamamoto}},
  \bibinfo {author} {\bibfnamefont {I.}~\bibnamefont {Danshita}}, \bibinfo
  {author} {\bibfnamefont {N.}~\bibnamefont {Yokoshi}}, \ and\ \bibinfo
  {author} {\bibfnamefont {S.}~\bibnamefont {Tsuchiya}},\ }\href {\doibase
  10.1103/PhysRevA.80.063627} {\bibfield  {journal} {\bibinfo  {journal} {Phys.
  Rev. A}\ }\textbf {\bibinfo {volume} {80}},\ \bibinfo {pages} {063627}
  (\bibinfo {year} {2009})}\BibitemShut {NoStop}%
\bibitem [{\citenamefont {Dzyaloshinsky}(1958)}]{Dzyaloshinsky1958}%
  \BibitemOpen
  \bibfield  {author} {\bibinfo {author} {\bibfnamefont {I.}~\bibnamefont
  {Dzyaloshinsky}},\ }\href {\doibase
  https://doi.org/10.1016/0022-3697(58)90076-3} {\bibfield  {journal} {\bibinfo
   {journal} {Journal of Physics and Chemistry of Solids}\ }\textbf {\bibinfo
  {volume} {4}},\ \bibinfo {pages} {241 } (\bibinfo {year} {1958})}\BibitemShut
  {NoStop}%
\bibitem [{\citenamefont {Moriya}(1960)}]{Moriya1960}%
  \BibitemOpen
  \bibfield  {author} {\bibinfo {author} {\bibfnamefont {T.}~\bibnamefont
  {Moriya}},\ }\href {\doibase 10.1103/PhysRev.120.91} {\bibfield  {journal}
  {\bibinfo  {journal} {Phys. Rev.}\ }\textbf {\bibinfo {volume} {120}},\
  \bibinfo {pages} {91} (\bibinfo {year} {1960})}\BibitemShut {NoStop}%
\bibitem [{\citenamefont {Jaksch}\ and\ \citenamefont
  {Zoller}(2003)}]{Jaksch2003}%
  \BibitemOpen
  \bibfield  {author} {\bibinfo {author} {\bibfnamefont {D.}~\bibnamefont
  {Jaksch}}\ and\ \bibinfo {author} {\bibfnamefont {P.}~\bibnamefont
  {Zoller}},\ }\href {\doibase 10.1088/1367-2630/5/1/356} {\bibfield  {journal}
  {\bibinfo  {journal} {New Journal of Physics}\ }\textbf {\bibinfo {volume}
  {5}},\ \bibinfo {pages} {56} (\bibinfo {year} {2003})}\BibitemShut {NoStop}%
\bibitem [{\citenamefont {Schweikhard}\ \emph {et~al.}(2004)\citenamefont
  {Schweikhard}, \citenamefont {Coddington}, \citenamefont {Engels},
  \citenamefont {Mogendorff},\ and\ \citenamefont {Cornell}}]{Schweikhard2004}%
  \BibitemOpen
  \bibfield  {author} {\bibinfo {author} {\bibfnamefont {V.}~\bibnamefont
  {Schweikhard}}, \bibinfo {author} {\bibfnamefont {I.}~\bibnamefont
  {Coddington}}, \bibinfo {author} {\bibfnamefont {P.}~\bibnamefont {Engels}},
  \bibinfo {author} {\bibfnamefont {V.}~\bibnamefont {Mogendorff}}, \ and\
  \bibinfo {author} {\bibfnamefont {E.~A.}\ \bibnamefont {Cornell}},\
  }\href@noop {} {\bibfield  {journal} {\bibinfo  {journal} {Physical review
  letters}\ }\textbf {\bibinfo {volume} {92}},\ \bibinfo {pages} {040404}
  (\bibinfo {year} {2004})}\BibitemShut {NoStop}%
\bibitem [{\citenamefont {Lin}\ \emph {et~al.}(2009)\citenamefont {Lin},
  \citenamefont {Compton}, \citenamefont {Jim{\'e}nez-Garc{\'\i}a},
  \citenamefont {Porto},\ and\ \citenamefont {Spielman}}]{Lin2009}%
  \BibitemOpen
  \bibfield  {author} {\bibinfo {author} {\bibfnamefont {Y.~J.}\ \bibnamefont
  {Lin}}, \bibinfo {author} {\bibfnamefont {R.~L.}\ \bibnamefont {Compton}},
  \bibinfo {author} {\bibfnamefont {K.}~\bibnamefont
  {Jim{\'e}nez-Garc{\'\i}a}}, \bibinfo {author} {\bibfnamefont {J.~V.}\
  \bibnamefont {Porto}}, \ and\ \bibinfo {author} {\bibfnamefont {I.~B.}\
  \bibnamefont {Spielman}},\ }\href {\doibase 10.1038/nature08609} {\bibfield
  {journal} {\bibinfo  {journal} {Nature}\ }\textbf {\bibinfo {volume} {462}},\
  \bibinfo {pages} {628} (\bibinfo {year} {2009})}\BibitemShut {NoStop}%
\bibitem [{\citenamefont {An}\ \emph {et~al.}(2017)\citenamefont {An},
  \citenamefont {Meier},\ and\ \citenamefont {Gadway}}]{An2017}%
  \BibitemOpen
  \bibfield  {author} {\bibinfo {author} {\bibfnamefont {F.~A.}\ \bibnamefont
  {An}}, \bibinfo {author} {\bibfnamefont {E.~J.}\ \bibnamefont {Meier}}, \
  and\ \bibinfo {author} {\bibfnamefont {B.}~\bibnamefont {Gadway}},\ }\href
  {\doibase 10.1126/sciadv.1602685} {\bibfield  {journal} {\bibinfo  {journal}
  {Science Advances}\ }\textbf {\bibinfo {volume} {3}} (\bibinfo {year}
  {2017}),\ 10.1126/sciadv.1602685},\ \Eprint
  {http://arxiv.org/abs/https://advances.sciencemag.org/content/3/4/e1602685.full.pdf}
  {https://advances.sciencemag.org/content/3/4/e1602685.full.pdf} \BibitemShut
  {NoStop}%
\bibitem [{\citenamefont {Aidelsburger}\ \emph {et~al.}(2013)\citenamefont
  {Aidelsburger}, \citenamefont {Atala}, \citenamefont {Lohse}, \citenamefont
  {Barreiro}, \citenamefont {Paredes},\ and\ \citenamefont
  {Bloch}}]{Aidelsburger2013}%
  \BibitemOpen
  \bibfield  {author} {\bibinfo {author} {\bibfnamefont {M.}~\bibnamefont
  {Aidelsburger}}, \bibinfo {author} {\bibfnamefont {M.}~\bibnamefont {Atala}},
  \bibinfo {author} {\bibfnamefont {M.}~\bibnamefont {Lohse}}, \bibinfo
  {author} {\bibfnamefont {J.~T.}\ \bibnamefont {Barreiro}}, \bibinfo {author}
  {\bibfnamefont {B.}~\bibnamefont {Paredes}}, \ and\ \bibinfo {author}
  {\bibfnamefont {I.}~\bibnamefont {Bloch}},\ }\href {\doibase
  10.1103/PhysRevLett.111.185301} {\bibfield  {journal} {\bibinfo  {journal}
  {Phys. Rev. Lett.}\ }\textbf {\bibinfo {volume} {111}},\ \bibinfo {pages}
  {185301} (\bibinfo {year} {2013})}\BibitemShut {NoStop}%
\bibitem [{\citenamefont {Miyake}\ \emph {et~al.}(2013)\citenamefont {Miyake},
  \citenamefont {Siviloglou}, \citenamefont {Kennedy}, \citenamefont {Burton},\
  and\ \citenamefont {Ketterle}}]{Miyake2013}%
  \BibitemOpen
  \bibfield  {author} {\bibinfo {author} {\bibfnamefont {H.}~\bibnamefont
  {Miyake}}, \bibinfo {author} {\bibfnamefont {G.~A.}\ \bibnamefont
  {Siviloglou}}, \bibinfo {author} {\bibfnamefont {C.~J.}\ \bibnamefont
  {Kennedy}}, \bibinfo {author} {\bibfnamefont {W.~C.}\ \bibnamefont {Burton}},
  \ and\ \bibinfo {author} {\bibfnamefont {W.}~\bibnamefont {Ketterle}},\
  }\href {\doibase 10.1103/PhysRevLett.111.185302} {\bibfield  {journal}
  {\bibinfo  {journal} {Phys. Rev. Lett.}\ }\textbf {\bibinfo {volume} {111}},\
  \bibinfo {pages} {185302} (\bibinfo {year} {2013})}\BibitemShut {NoStop}%
\bibitem [{\citenamefont {Dirac}(1931)}]{Dirac1931}%
  \BibitemOpen
  \bibfield  {author} {\bibinfo {author} {\bibfnamefont {P.~A.~M.}\
  \bibnamefont {Dirac}},\ }\href {\doibase 10.1098/rspa.1931.0130} {\bibfield
  {journal} {\bibinfo  {journal} {Proceedings of the Royal Society of London A:
  Mathematical, Physical and Engineering Sciences}\ }\textbf {\bibinfo {volume}
  {133}},\ \bibinfo {pages} {60} (\bibinfo {year} {1931})}\BibitemShut
  {NoStop}%
\bibitem [{\citenamefont {{Wu}}\ and\ \citenamefont {{Yang}}(1976)}]{Wu1976}%
  \BibitemOpen
  \bibfield  {author} {\bibinfo {author} {\bibfnamefont {T.~T.}\ \bibnamefont
  {{Wu}}}\ and\ \bibinfo {author} {\bibfnamefont {C.~N.}\ \bibnamefont
  {{Yang}}},\ }\href {\doibase 10.1016/0550-3213(76)90143-7} {\bibfield
  {journal} {\bibinfo  {journal} {Nuclear Physics B}\ }\textbf {\bibinfo
  {volume} {107}},\ \bibinfo {pages} {365} (\bibinfo {year}
  {1976})}\BibitemShut {NoStop}%
\bibitem [{\citenamefont {Baxter}(2007)}]{Baxter2007}%
  \BibitemOpen
  \bibfield  {author} {\bibinfo {author} {\bibfnamefont {R.}~\bibnamefont
  {Baxter}},\ }\href {https://books.google.ca/books?id=G3owDULfBuEC} {\emph
  {\bibinfo {title} {Exactly Solved Models in Statistical Mechanics}}},\ Dover
  books on physics\ (\bibinfo  {publisher} {Dover Publications},\ \bibinfo
  {year} {2007})\BibitemShut {NoStop}%
\bibitem [{\citenamefont {Chandra}\ \emph {et~al.}(1990)\citenamefont
  {Chandra}, \citenamefont {Coleman},\ and\ \citenamefont
  {Larkin}}]{Chandra1990}%
  \BibitemOpen
  \bibfield  {author} {\bibinfo {author} {\bibfnamefont {P.}~\bibnamefont
  {Chandra}}, \bibinfo {author} {\bibfnamefont {P.}~\bibnamefont {Coleman}}, \
  and\ \bibinfo {author} {\bibfnamefont {A.~I.}\ \bibnamefont {Larkin}},\
  }\href {\doibase 10.1103/PhysRevLett.64.88} {\bibfield  {journal} {\bibinfo
  {journal} {Phys. Rev. Lett.}\ }\textbf {\bibinfo {volume} {64}},\ \bibinfo
  {pages} {88} (\bibinfo {year} {1990})}\BibitemShut {NoStop}%
\bibitem [{\citenamefont {Miyashita}\ and\ \citenamefont
  {Shiba}(1984)}]{Miyashita1984}%
  \BibitemOpen
  \bibfield  {author} {\bibinfo {author} {\bibfnamefont {S.}~\bibnamefont
  {Miyashita}}\ and\ \bibinfo {author} {\bibfnamefont {H.}~\bibnamefont
  {Shiba}},\ }\href {\doibase 10.1143/JPSJ.53.1145} {\bibfield  {journal}
  {\bibinfo  {journal} {Journal of the Physical Society of Japan}\ }\textbf
  {\bibinfo {volume} {53}},\ \bibinfo {pages} {1145} (\bibinfo {year}
  {1984})},\ \Eprint
  {http://arxiv.org/abs/https://doi.org/10.1143/JPSJ.53.1145}
  {https://doi.org/10.1143/JPSJ.53.1145} \BibitemShut {NoStop}%
\end{thebibliography}%

\end{document}